\documentclass[a4paper,11pt]{article}
\pdfoutput=1 % if your are submitting a pdflatex (i.e. if you have
\usepackage{jcappub} % for details on the use of the package, please
\usepackage[T1]{fontenc} % if needed
\usepackage{booktabs}
\graphicspath{{plots/}}
\title{Strong Bounds on Sum of Neutrino Masses in a 12 Parameter Extended Scenario with Non-Phantom Dynamical Dark Energy ($w(z)\geq -1$)}

\author[a,b,1]{Shouvik Roy Choudhury,\note{Corresponding author.}}
\author[c]{Abhishek Naskar}

\affiliation[a]{Harish-Chandra Research Institute\\Chhatnag Road, Jhunsi, Prayagraj (Allahabad) - 211019, India}
\affiliation[b]{Homi Bhabha National Institute\\ Training School Complex, Anushaktinagar, Mumbai - 400094, India}
\affiliation[c]{Indian Statistical Institute, Kolkata \\ 203 BT Road, Kolkata-700108, India}

\emailAdd{shouvikroychoudhury@hri.res.in}
\emailAdd{abhiatrkmrc@gmail.com}

\abstract{We obtained constraints on a 12 parameter extended cosmological scenario including non-phantom dynamical dark energy (NPDDE) with CPL parametrization. We also include the six $\Lambda$CDM parameters, number of relativistic neutrino species ($N_{\textrm{eff}}$) and sum over active neutrino masses ($\sum m_{\nu}$), tensor-to-scalar ratio ($r_{0.05}$), and running of the spectral index ($n_{run}$). We use CMB Data from Planck 2015; BAO Measurements from SDSS BOSS DR12, MGS, and 6dFS; SNe Ia Luminosity Distance measurements from the Pantheon Sample; CMB B-mode polarization data from BICEP2/Keck collaboration (BK14); Planck lensing data; and a prior on Hubble constant ($73.24\pm1.74$ km/sec/Mpc) from local measurements (HST). We have found strong bounds on the sum of the active neutrino masses. For instance, a strong bound of $\sum m_{\nu} <$ 0.123 eV (95\% C.L.) comes from Planck+BK14+BAO. Although we are in such an extended parameter space, this bound is stronger than a bound of $\sum m_{\nu} <$ 0.158 eV (95\% C.L.) obtained in $\Lambda \textrm{CDM}+\sum m_{\nu}$ with Planck+BAO. Varying $A_{\textrm{lens}}$ instead of $r_{0.05}$ however leads to weaker bounds on $\sum m_{\nu}$. Inclusion of the HST leads to the standard value of $N_{\textrm{eff}} = 3.045$ being discarded at more than 68\% C.L., which increases to 95\% C.L. when we vary $A_{\textrm{lens}}$ instead of $r_{0.05}$, implying a small preference for dark radiation, driven by the $H_0$ tension.}

\begin{document}
\maketitle
\flushbottom

\section{Introduction}
\label{sec:1}

Recent observations suggest that the universe can be modelled according to the six parameter $\Lambda$CDM model, where structure formation is explained by cold dark matter physics (CDM) and recent acceleration of the universe is explained by vacuum energy $\Lambda$ which is the candidate for dark energy. There are however possible extensions to the standard $\Lambda$CDM. Cosmic neutrino background and Inflationary Gravitational waves (IGWs/tensors) are theoretically well motivated. Among them, cosmic neutrino background ($\rm C\nu \rm B$) is indirectly confirmed by the CMB measurements of the Planck satellite, where the current preferred value of the effective number of extra radiation species at recombination, $N_{\textrm {eff}} = 2.92^{+0.36}_{-0.37}$ (95\%, TT,TE,EE+lowE) \cite{Aghanim:2018eyx} in a minimal $\Lambda \textrm{CDM}+ N_{\textrm {eff}}$, is very far away away from the value of $N_{\textrm{eff}} = 0$. The theoretically predicted value of $N_{\textrm {eff}} = 3.045$ \cite{deSalas:2016ztq} considering three active neutrinos as the only relativistic species apart from photons during recombination, is completely compatible with this bound, implying consistency with $\Lambda$CDM. In standard model of particle physics, neutrinos are massless. But terrestrial neutrino oscillation experiments have strongly confirmed that neutrinos have small masses. While strongest upper bounds on the sum of masses of the three active neutrino mass eigenstates $\sum m_{\nu}$) come from cosmology, it is still unable to provide any lower bound, indicating that the standard model assumption of $\sum m_{\nu} = 0$ is consistent with current data. For instance, Planck 2018 results \cite{Aghanim:2018eyx} provided a bound of $\sum m_{\nu} < 0.12$ eV (95\% C.L.) in the minimal $\Lambda CDM+\sum m_{\nu}$ for the TT,TE,EE+ lowE + lensing + BAO data combination. There are numerous other analyses with different datasets which provide bounds of $\sum m_{\nu} \lesssim 0.15$ eV (95\% C.L.) \cite{Capozzi:2017ipn,Palanque-Delabrouille:2015pga,DiValentino:2015wba,Cuesta:2015iho,Huang:2015wrx,Moresco:2016nqq,Giusarma:2016phn,Vagnozzi:2017ovm,Couchot:2017pvz,Caldwell:2017mqu,Doux:2017tsv,Wang:2017htc,Chen:2017ayg,Upadhye:2017hdl,Salvati:2017rsn,Nunes:2017xon,Zennaro:2017qnp, Wang:2018lun,Choudhury:2018byy}, i.e, cosmological data is becoming more and more effective in constraining neutrino masses. Please also see \cite{Lattanzi:2017ubx,deSalas:2018bym}, which provide detailed reviews on current status and future prospects of constraining neutrino masses and determining their ordering from cosmology and other data. 

Again, while $\rm C\nu \rm B$ is indirectly detected, existence of IGWs is still to be confirmed. The main probe for IGWs is the CMB B-mode polarization, and the corresponding important parameter is the tensor-to-scalar ratio ($r$). The currently available observations can only put an upper bound on the tensor to scalar ratio: $r_{0.05} < 0.06$ (95\% C.L.; at a pivot scale of $k_* = 0.05h Mpc^{-1}$) \cite{Ade:2018gkx}, implying that $r=0$ is consistent with current data. While $\Lambda$CDM has its success there are also parameter tensions between CMB and non-CMB data within the $\Lambda$CDM  model. One of the most important limitations of $\Lambda$CDM is that high redshift (CMB) and low redshift (local universe) measurements gives different values of Hubble constant. The Planck 2018 results \cite{Aghanim:2018eyx} provide $H_0=67.27\pm 0.60$ km/sec/Mpc (68\% C.L.) for TT,TE,EE + lowE in $\Lambda$CDM (with $\sum m_{\nu}$ fixed at 0.06 eV), and recent direct measurement gives $H_0=73.24\pm 1.74$ km/sec/Mpc (68\% C.L., hereafter HST) \cite{Riess:2016jrr}. There is roughly a $3\sigma$ inconsistency between these datasets. Recent strong lensing observations from the H0LiCOW program \cite{Bonvin:2016crt} provides $H_0=71.9^{+2.4}_{-3.0}$ km/sec/Mpc (68\% C.L.) and partially confirms the tension. CMB data also has $\sim 2\sigma$ tensions in the measurements of $\Omega_{m}$ and $\sigma_8$ with x-ray galaxy cluster measurements \cite{Bohringer:2014ooa} or cosmic shear surveys like CFHTLenS \cite{Heymans:2012gg} and KiDS-450 \cite{Hildebrandt:2016iqg}. For instance, the KiDS-450 survey measures a combined quantity $S_8 \equiv \sigma_8 \sqrt{\Omega_m/0.3} = 0.745 \pm 0.039$ (68\% C.L.) which has a more than 2$\sigma$ tension with Planck 2018, which prefers a much higher value of $S_8 = 0.834 \pm 0.016$ (68\% C.L.; TT,TE,EE + lowE).

Apart from inconsistencies among high and low redshift datasets, there are several internal inconsistencies in the Planck data itself. Parameter estimations in $\Lambda$CDM differ when considering small scale ($l\geq 1000$) and high or intermediate scale ($l<1000$) temperature data separately \cite{Addison:2015wyg}. This is especially true for the measured value of $H_0$ which is much lower when obtained from the $l\geq 1000$ data than when obtained from the $l<1000$ data. 	
%Moreover, the measured value of the optical depth to reionization $\tau = 0.055 \pm 0.009$ by the 2016 Planck re-analysis of its high frequency channels in low-$l$ ($l$<29) EE data \cite{Aghanim:2016yuo} is at 1.7$\sigma$ tension with the measured value of $\tau = 0.078 \pm 0.024$ from Planck TT+lowP \cite{Ade:2015xua}. 
Another puzzling inconsistency in $\Lambda$CDM with Planck data is that the latest measurement of lensing parameter by Planck 2018, $A_{\textrm{lens}}=1.180\pm0.065$ (68\% C.L.) in a $\Lambda \textrm{CDM} + A_{\textrm{lens}}$ model \cite{Aghanim:2018eyx} is 2.8$\sigma$ level higher than $\Lambda$CDM prediction of $A_{\textrm{lens}}=1$. See also \cite{Calabrese:2008rt,Renzi:2017cbg} on the $A_{\textrm{lens}}$ problem. 

A possible explanation for these tensions is the systematics of the observations. But it is also possible that we need physics beyond $\Lambda$CDM and standard model of particle physics. These inconsistencies in $\Lambda \textrm{CDM}$ model and different datasets have motivated several studies of cosmological scenarios in extended parameter spaces \cite{DiValentino:2015ola,DiValentino:2016hlg,DiValentino:2017zyq,Ade:2015rim,Pourtsidou:2016ico,Ballardini:2016cvy,Grandis:2016fwl,Zhao:2017urm,Yang:2017amu,Kumar:2016zpg,Joudaki:2016kym,Karwal:2016vyq,Ko:2016uft,Bernal:2016gxb,Archidiacono:2016kkh,Qing-Guo:2016ykt,Chacko:2016kgg,Zhang:2017idq,Zhao:2017cud,Sola:2017jbl,Brust:2017nmv,Vagnozzi:2017ovm,Vagnozzi:2018jhn,Lambiase:2018ows,Song:2018zyl,Bhattacharyya:2018fwb,Martins:2018bzo,Choudhury:2018byy,Xia:2016vnp, Wang:2016tsz, Kumar:2017dnp}. 
Recent studies have also analyzed models with as large as twelve parameters \cite{DiValentino:2015ola,DiValentino:2016hlg,DiValentino:2017zyq}. The motivation behind studying such a large parameter space is that $\Lambda \textrm{CDM}$ currently seems to be an over-simplification. Indeed, there is no reason to fix $\sum m_{\nu}$ to 0.06 eV (95\% C.L.), since it is only approximately the minimum sum of masses required for normal hierarchy of neutrinos and this mass might not be an accurate one. Massive neutrinos produce distinct effects on CMB and large scale structure data and this has been widely studied \cite{LESGOURGUES2006307,doi:10.1146/annurev-nucl-102010-130252,Lesgourgues-Pastor,ABAZAJIAN201566,1367-2630-16-6-065002,1475-7516-2017-02-052}. Again, the discrepancy with Planck and HST might be explained by a dark radiation species contributing to $N_{\textrm{eff}}$ \cite{Riess:2016jrr}. Similarly, existence of tensor perturbations are theoretically well motivated and there seem no reason to not to include them in a analysis. 

Apart from massive neutrinos and tensors, another extension to  $\Lambda$CDM which has recently received a lot of attention is dynamical dark energy, where the dark energy (DE) equation of state (EoS) is not fixed at $w=-1$ or some other constant, but is varying with time \cite{Zhao:2017cud}. Dark energy is one of the biggest puzzles, not only in Cosmology, but in the whole of Physics. Currently available datasets, in this era of precision cosmology, can provide us with much better bounds on DE equation of state than it was previously possible. Thus it seems simplistic and unnecessary to assume dark energy as just a cosmological constant, especially when from the quantum field theoretic point of view, it has been a very difficult thing to explain \cite{Padilla:2015aaa}. Hence, in this work, with massive neutrinos, tensors, and dynamical dark energy included, we consider a largely extended cosmology compared to a standard one. 

However, we do not include the full dynamical dark energy range. The $w=-1$ line divides the dynamics of dark energy in two distinct regions, phantom ($w<-1$) and non-phantom ($w\geq-1$). In this work, we discard the phantom region as first done in \cite{Vagnozzi:2018jhn} in the context of cosmological neutrino mass constraints, and specifically consider a non-phantom scenario, since in a universe with phantom dark energy ($w<-1$), the dark energy density reaches infinity in a finite time leading to dissociation of all bound states, i.e., the so called Big Rip, and seems unphysical in that sense \cite{Caldwell:2003vq,Vikman:2004dc}. From field theory perspective, Dark energy models with a single scalar field are not able to go across the $w=-1$ line (i.e., the phantom barrier) and more general models that allow it demand extra degrees of freedom to supply stability gravitationally \cite{Fang:2008sn}. Phantom dark energy accommodating field theories are usually plagued with one or more of the following problems like  Lorentz violation, unstable vacuum, superluminal modes, ghosts, non-locality, or instability to quantum corrections. On the other hand, however, it is possible to make theories free of such abnormalities by using effects like photon-axion conversion or modified gravity  which leads to an apparent $w<-1$ (see \cite{Ludwick:2017tox} for a brief review), or vacuum phase transition (\cite{DiValentino:2017rcr}), which produces a phantom behaviour of the DE EoS. Nonetheless, there are single scalar field theories like quintessence \cite{Linder:2007wa,Caldwell:2005tm,Zlatev:1998tr} which are relatively well motivated theoretically, and are non-phantom in nature. So, in this work we limit ourselves to such theories. Our main motivation to do this work has been to study how effective the currently available datasets are in constraining the cosmological parameters (especially the sum of neutrino masses) in a non-phantom dynamical dark energy scenario instead of a cosmological constant, with minimal assumptions about other parameters coming from the massive neutrinos and tensor sector.

In this work we have first considered a 12 parameter extended scenario with 6 usual $\Lambda \textrm{CDM}$ parameters, two dynamical dark energy parameters ($w_0-w_a$ approach, CPL parametrization) with $w(z)\geq -1$, two neutrino parameters ($N_{\textrm{eff}}$ and $\sum m_{\nu}$), and two inflationary parameters ($r_{0.05}$ and the running of the spectral index, $n_{run} \equiv dn_s/d\textrm{ln~k}$). We performed a Bayesian analysis to constrain parameters using different combinations of latest available datasets: (1) Cosmic Microwave Background temperature and polarization data from Planck 2015; (2)the latest data released from the BICEP2/Keck Collaboration for the BB mode of the CMB spectrum (BK14); (3) Baryon Acoustic Oscillation Measurements from SDSS III BOSS DR12, MGS and 6dFGS; (4) Supernovae Type Ia Luminosity Distance Measurements from the newly released Pantheon Sample, (5) Planck 2015 lensing data; and (6) the HST Gaussian prior ($H_0=73.24\pm 1.74$ km/sec/Mpc (68\% C.L.)) on Hubble constant. Next we turned off the tensor perturbations (i.e., removed $r_{0.05}$) and constrained this 11 parameter scenario with the same datasets except BK14. Finally we add a new parameter $A_{\textrm{lens}}$ and again constrain this 12 parameter expended space with the mentioned datasets. We emphasize here that this is the first time someone has evaluated the non-phantom dark energy scenario in a 12 parameter extended space. Our main focus in this paper is on sum of neutrino masses, however we provide the constraints on all the varying parameters. Here we would also like to emphasize that we take the datasets at face value, i.e., any discrepancy or tension between datasets in our model is assumed to have a physical reason and not due to unknown systematics involved in the experiments. Also, it is imperative to point out that the best bounds on sum of neutrino masses that we have presented, are strong and comparable or better to the bounds provided by the recently released Planck 2018 results \cite{Aghanim:2018eyx} in the $\Lambda \textrm{CDM}+\sum m_{\nu}$ model. Hence our results remain very much relevant although we have used the Planck 2015 data. 

It is imperative that we also mention three recent papers which have helped in building the motivation for this work, and also the difference in our analyses with the said papers. In \cite{DiValentino:2017zyq}, the authors constrained the dark energy dynamics in an extended 12 parameter model, but they included both the phantom and non-phantom sectors of dark energy,and did not consider any tensor modes. In our analysis, we also use 12 parameters, but we have included tensor perturbations, use newer datasets, and more importantly, we have discarded the phantom DE sector as explained above. We would like to mention that this does affect the bounds on $\sum m_{\nu}$ greatly, i.e., they become far stronger compared to the case where phantom DE is included. Bounds on other cosmological parameters also improve. The fact that the neutrino mass bounds from cosmology improve greatly in a non-phantom dark energy scenario, and are stronger even compared to the minimal $\Lambda\textrm{CDM}+\sum m_{\nu}$ case was shown by two recent papers \cite{Vagnozzi:2018jhn,Choudhury:2018byy}. However, analyses in both of these papers were done in smaller parameter spaces, and none of these two papers have $N_{\textrm{eff}}$ and $A_{\textrm{lens}}$ as free parameters as we have. Consequently, they have not touched the issues like the possibility of extra radiation species and the $A_{\textrm{lens}}$ problem. Ref. \cite{Vagnozzi:2018jhn} also uses older datasets. In this paper, we have, for the first time, shown that neutrino mass bounds can indeed be stronger than the minimal $\Lambda \textrm{CDM} + \sum m_{\nu}$ model even in a 12 parameter extended scenario if one considers non-phantom dark energy, even though one expects the bounds to relax in such a large extended space. We have also shown that it is possible to effectively constrain cosmological parameters with some reasonable 1-$\sigma$ ranges with current cosmological data, in a 12 parameter expended scenario with non-phantom dark energy.   

This paper is arranged as follows: in section \ref{sec:2} we describe the cosmological models used in this paper and the prior ranges of parameters used, along with a brief description of the CPL parametrization. In section \ref{sec:3}
we briefly describe the datasets used in this work. In section \ref{sec:4} we present our analysis results. In section \ref{sec:6}, we further discuss how the neutrino mass bounds will change in the three models with new values of $\tau$ and $A_{\textrm{lens}}$ obtained by the new Planck 2018 collaboration \cite{Aghanim:2018eyx}. We provide a discussion and summary in section \ref{sec:5}. The main results are in tables \ref{table:2}, \ref{table:3}, and \ref{table:4}.

\section{Models}
\label{sec:2}

In this work we have considered 3 different cosmological scenarios to obtain bounds on the cosmological parameters. Below we list the vector of parameters to vary in each of these cosmological scenarios.

For NPDDE11+$r$ model with 12 parameters: 
\begin{equation}\label{eqn:1}
\theta \equiv \left[\omega_c, ~\omega_b, ~\Theta_s,~\tau, ~n_s, ~ln [ 10^{10} A_s],  w_0, w_a, N_{\textrm{eff}}, \sum m_{\nu}, r_{0.05}, n_{run} \right].
\end{equation} 

For NPDDE11 model with 11 parameters:
\begin{equation}\label{eqn:2}
\theta \equiv \left[\omega_c, ~\omega_b, ~\Theta_s,~\tau, ~n_s, ~ln [ 10^{10} A_s], w_0, w_a, N_{\textrm{eff}}, \sum m_{\nu}, n_{run} \right].
\end{equation}

For NPDDE11+$A_{\textrm{lens}}$ model with 12 parameters :
\begin{equation}\label{eqn:3}
\theta \equiv \left[\omega_c, ~\omega_b, ~\Theta_s,~\tau, ~n_s, ~ln [ 10^{10} A_s], w_0, w_a, N_{\textrm{eff}}, \sum m_{\nu}, n_{run}, A_{\textrm{lens}} \right].
\end{equation}

In this analysis, the first model, NPDDE11+$r$, comprises of six additional parameters on top of $\Lambda$CDM model. The six parameters of $\Lambda$CDM are: present day cold dark matter energy density $\omega_c \equiv \Omega_c h^2$, present day baryon energy density $\omega_b \equiv \Omega_b h^2$,  reionization optical depth $\tau$, spectral tilt and amplitude of primordial scalar power spectrum $n_s$ and $A_s$ (evaluated at pivot scale $k_*=0.05 h Mpc^{-1}$) and $\Theta_s$ is the ratio between the sound horizon and the angular diameter distance at decoupling.. For our analysis we are adding the following parameters: two dark energy parameters $w_0$ and $w_a$, effective number of relativistic species at recombination $N_{\textrm{eff}}$, total neutrino mass $\sum m_{\nu}$, the tensor-to-scalar ratio $r_{0.05}$ (evaluated at pivot scale $k_*=0.05 h Mpc^{-1}$) and the running of spectral index of primordial power spectrum $n_{run}$($\equiv dn_s/d\textrm{ln~k}$). In this model, the gravitational lensing amplitude of the CMB angular spectra $A_{\textrm{lens}}$ is fixed at the $\Lambda$CDM predicted value of unity.

We also consider two other scenarios. In the NPDDE11 model, we do not run the tensor perturbations and constrain the parameter space considering scalar only perturbations. In the NPDDE11+$A_{\textrm{lens}}$ model we also allow the $A_{\textrm{lens}}$ parameter to vary. This is since the cause of the  $A_{\textrm{lens}}$-anomaly is unknown and therefore it is important to look into the effect of varying $A_{\textrm{lens}}$ on the constraints of rest of the parameter space. 

\emph{CPL Parametrization:}
For dark energy dynamics we use the famous Chevallier-Polarski-Linder (CPL) parametrization \cite{doi:10.1142/S0218271801000822,PhysRevLett.90.091301} which uses a varying equation of state in terms of the redshift $z$ and two parameters $w_0$ and $w_a$:
\begin{equation}
w (z) \equiv w_0 + w_a (1 - a) = w_0 + w_a \frac{z}{1+z}.
\end{equation}
This uses the Taylor expansion of the equation of state in powers of the scale factor $a = 1/(1+z)$ and takes only the first two terms. Here $w(z=0) = w_0$ is the dark energy EoS at present day ($z=0$), whereas $w(z\rightarrow \infty) = w_0+w_a$ is the dark energy EoS in the beginning of the universe; and $w(z)$ is a monotonic function between these two times. Therefore, to constrain only the NPDDE region of the parameter space i.e. $w(z) \geq -1$ it is enough to apply these hard priors:
\begin{equation}\label{eqn:4}
w_0 \geq -1;~~~~~~~w_0+w_a\geq -1.
\end{equation}

For the cosmological parameters mentioned in eqs.~\ref{eqn:1}--\ref{eqn:3}, we have assumed flat priors which are listed in table~\ref{table:1}, along with hard priors given in eq.~\ref{eqn:4}. We obtain the constraints using the Markov Chain Monte Carlo (MCMC) sampler CosmoMC \cite{PhysRevD.66.103511} which uses CAMB \cite{Lewis:1999bs} as the Boltzmann code and the Gelman and Rubin statistics \cite{doi:10.1080/10618600.1998.10474787} to estimate the convergence of chains. All our chains reached the convergence criterion of $R-1<0.01$.

\begin{table}
\begin{center}
\begin{tabular}{c c}
\hline
Parameter                    & Prior\\
\hline
$\Omega_{\rm b} h^2$         & [0.005,0.1]\\
$\Omega_{\rm c} h^2$         & [0.001,0.99]\\
$\tau$                       & [0.01,0.8]\\
$n_s$                        & [0.8, 1.2]\\
$\log[10^{10}A_{s}]$         & [2,4]\\
$\Theta_{\rm s}$             & [0.5,10]\\ 
$w_0$                        & [-1,-0.33]\\
$w_a$                        & [-2,2]\\
$N_{\rm eff}$                & [0.05,10]\\ 
$\sum m_\nu$ (eV)            & [0,5]\\
$r_{0.05}$                   & [0,1] \\
$n_{run}$                    & [-1,1]\\
$A_{\textrm{lens}}$          & [0,10]\\
\hline
\end{tabular}
\end{center}
\caption{\label{table:1} Flat priors on the main cosmological parameters constrained in this paper.}
\label{priors}
\end{table}

\section{Datasets}
\label{sec:3}
Below, we provide a description of the datasets used in our analyses. We have used different combinations of these datasets. 

\emph{Cosmic Microwave Background: Planck 2015}:

We have used measurements of the CMB temperature, polarization, and temperature-polarization cross-correlation spectra from the Planck 2015 data release \cite{Planck2015-I,Ade:2015xua}. We use a combination of the high-$l$ (30 $\leq$ $l$ $\leq$ 2508) and low-$l$ (2 $\leq$ $l$ $\leq$ 29) TT likelihood. Along with that, we also include the high-$l$ (30 $\leq$ $l$ $\leq$ 1996) EE and TE likelihood and the low-$l$ (2 $\leq$ $l$ $\leq$ 29) polarization likelihood. We refer to this whole dataset as Planck.  \\

\emph{Baryon Acoustic Oscillations (BAO) Measurements}:

We use measurements of the BAO signal obtained from different galaxy surveys in this work. We include the SDSS-III BOSS DR12 Consensus sample (\cite{doi:10.1093/mnras/stx721} which includes LOWZ and CMASS galaxy samples at $z_{\textrm{eff}} =$ 0.38, 0.51 and 0.61). Along with it, we also include the DR7 MGS at $z_{\textrm{eff}} = 0.15$ \cite{doi:10.1093/mnras/stv154}, and the 6dFGS survey at $z_{\textrm{eff}} = 0.106$ \cite{doi:10.1111/j.1365-2966.2011.19250.x}. We denote this full combination as BAO. Here $z_{\textrm{eff}}$ is the effective redshift of a survey.  
\\

\emph{Luminosity Distance Measurements from Type Ia Supernovae (SNe Ia)}: 

We also use Supernovae Type-Ia (SNe Ia) luminosity distance measurements from the Pantheon Sample \cite{Scolnic:2018rjj}. It comprises of data from  279 Pan-STARRS1 (PS1) Medium Deep Survey SNe Ia ($0.03 < z < 0.68$) and distance estimates of SNe Ia from SDSS, SNLS, various low-z and HST samples. This combined sample comprises of data from a total of 1048 SNe Ia with a redshift range of $0.01 < z < 2.3$ and is the largest one till date. We refer to this data as PAN from now on. This dataset supersedes the Joint Light-curve Analysis (JLA) sample which comprises of information from 740 spectroscopically confirmed SNe Ia \cite{JLA}. 
\\

\emph{BB Mode Spectrum of CMB}: 

We use the latest data available from BICEP2/Keck collaboration for the B mode polarization of CMB, which includes all data (range: $20 < l < 330$) taken up to and including 2014 \cite{Array:2015xqh}. This dataset is denoted as BK14. 
\\

\emph{Hubble Parameter Measurements}: 

We use a Gaussian prior of $73.24 \pm 1.74$ km/sec/Mpc (68\% C.L.) on $H_0$. This result is a recent 2.4\% determination of the local value of the Hubble parameter by \cite{Riess:2016jrr} which combines the anchor NGC 4258, Milky Way and LMC Cepheids. We denote this prior as HST. \\

While we use HST in most cases, we also provide some results with a prior with a lower value of $H_0 = 71.6\pm2.7$ km/sec/Mpc, which is based on the determination of the Hubble constant from the H0LiCOW programme \cite{Bonvin:2016crt}.We call this prior $H071p6$. This is to compare what happens when we use a $H_0$ prior that has less  tension with Planck than HST.\\

\emph{Planck Lensing Measurements}:

We also use the lensing potential measurements via reconstruction through the four point functions of Planck 2015   measurements of CMB~\cite{Ade:2015xua}. We simply refer to this data as lensing. 
\\

%\emph{Weak Lensing Measurements from CFHTLenS}:
% 
%We include the weak gravitational lensing data from the Canada-France-Hawaii Telescope Lensing Survey (CFHTLenS) Survey \cite{Heymans:2012gg} with conservative cuts was described in \cite{Ade:2015xua}. We refer to this dataset as WL.

\section{Results}
\label{sec:4}
We have split the results in the three smaller sections for the three different models we have studied. The description of models and datasets are given at section \ref{sec:2} and section \ref{sec:3} respectively. We have presented the results in the following order: first the NPDDE11+$r$ model, then the NPDDE11 model and lastly the NPDDE11+$A_{\textrm{lens}}$ model. All the marginalized limits quoted in the text or tables are at 68\% C.L. whereas upper limits are quoted at 95\% C.L.

\subsection{NPDDE11+$r$ model}
\label{sec:4:1}

\begin{table}
	
	\centering
%	\small\addtolength{\tabcolsep}{-4pt}
    \resizebox{\textwidth}{!}{
	\begin{tabular}{ccccccccc}
		\toprule
		\toprule
		 \vspace{0.2cm}
		Parameter &    Planck+BK14  &    Planck+BK14  &    Planck+BK14 &    Planck+BK14  &    Planck+BK14  &    Planck+BK14  &    Planck+BK14  \\\vspace{0.2cm}
		  &    +BAO  &   +BAO+PAN &    +BAO+HST+PAN &    +BAO+HST  &    +HST  &    +HST+PAN  &    +HST+lensing  \\
		\midrule
		\hspace{1mm}
		\vspace{ 0.2cm}
		 $\Omega_b h^2$  &    $0.02243\pm0.00021$  &    $0.02244\pm0.00020$  &    $0.02265\pm0.00019$  &    $0.02266\pm0.00019$  &    $0.02265\pm0.00020$  &    $0.02267\pm0.00020$  &    $0.02262\pm0.00020$   \\ \vspace{ 0.2cm}
		
		  $\Omega_c h^2$  &    $0.1189_{-0.0034}^{+0.0033}$  &    $0.1190\pm 0.0033$  &    $0.1231\pm0.0030$  &    $0.1233\pm 0.0031$  &    $0.1233\pm0.0030$  &    $0.1231\pm0.0031$  &    $0.1228_{-0.0032}^{+0.0029}$  \\ \vspace{ 0.2cm}
		
	 	$\tau$  &    $0.096^{+0.017}_{-0.018}$  &    $0.095^{+0.017}_{-0.018}$  &    $0.099^{+0.017}_{-0.018}$  &    $0.099 \pm 0.017$  &      $0.099 \pm 0.018$  &    $0.099 \pm 0.018$  &    $0.079\pm 0.015$  \\ \vspace{ 0.2cm}
		
	 	$n_s$  &    $0.969\pm0.010$  &    $0.969\pm0.009$  &    $0.981\pm0.009$  &    $0.982\pm 0.008$  &    $0.981\pm0.008$  &    $0.982\pm 0.008$  &    $0.982\pm0.009$  \\ \vspace{ 0.2cm}
		
	 	${\rm{ln}}(10^{10} A_s)$  &    $3.126_{-0.037}^{+0.036}$  &    $3.125_{-0.037}^{+0.036}$  &    $3.142^{+0.035}_{-0.036}$  &    $3.142^{+0.035}_{-0.036}$  &     $3.142\pm 0.036$  &    $3.143\pm 0.036$  &    $3.099\pm 0.030$  \\ \vspace{ 0.2cm}
		
	 	$\Theta_s$  &    $1.041\pm0.0005$  &    $1.041\pm0.0005$  &    $1.040\pm0.0004$  &    $1.040\pm0.0004$  &    $1.040\pm0.0004$  &    $1.040\pm0.0004$  &    $1.041\pm0.0004$  \\ \vspace{ 0.2cm}
	
	 	$r_{0.05}$   &    $<0.075$  &    $<0.074$  &    $<0.072$  &    $<0.071$  &    $<0.070$  &    $<0.071$  &    $<0.075$  \\ \vspace{ 0.2cm}	
	
	 	$H_0$ (km/s/Mpc)  &    $66.64^{+1.38}_{-1.37}$  &    $67.37^{+1.26}_{-1.25}$  &    $69.40\pm 1.05$  &     $69.13_{-1.08}^{+1.09}$   &    $69.14^{+1.36}_{-1.35}$  &    $69.57\pm 1.24$  &    $69.15\pm 1.38$  \\ \vspace{ 0.2cm}
		
	 	$\sigma_8$  &    $0.827\pm0.018$  &    $0.833\pm0.018$  &    $0.850\pm 0.017$  &    $0.847\pm 0.017$  &    $0.847\pm 0.018$  &    $0.851\pm 0.017$  &    $0.825\pm 0.015$  \\ \vspace{ 0.2cm}
		
	 	$\sum m_\nu$ (eV)  &    $<0.123$  &    $<0.126$  &    $<0.128$  &    $<0.129$  &    $<0.143$  &    $<0.132$  &    $<0.186$  \\ \vspace{ 0.2cm}
		
	 	$w_0$  &    $<-0.859$  &    $<-0.933$  &    $<-0.943$  &    $<-0.908$  &    $<-0.915$  &    $<-0.944$  &    $<-0.914$ \\ \vspace{ 0.2cm}
		
	 	$w_a$  &    $0.013^{+0.065}_{-0.077}$  &    $0.033^{+0.036}_{-0.063}$  &    $0.034_{-0.059}^{+0.031}$  &    $0.028_{-0.065}^{+0.046}$  &    $0.031_{-0.064}^{+0.041}$  &    $0.032_{-0.056}^{+0.029}$  &    $0.035_{-0.070}^{+0.043}$  \\ \vspace{ 0.2cm}
		
	 	$N_{\textrm{eff}}$  &    $3.082_{-0.211}^{+0.209}$  &    $3.089\pm 0.208$  &    $3.382\pm 0.181$  &    $3.392_{-0.186}^{+0.188}$  &    $3.391 \pm 0.185$  &    $3.390_{-0.185}^{+0.186}$  &    $3.393_{-0.197}^{+0.181}$  \\ \vspace{ 0.2cm}
		
	 	$n_{\rm run}$  &    $-0.00756^{+0.00793}_{-0.00797}$  &    $-0.00743^{+0.00811}_{-0.00815}$  &    $-0.00253_{-0.00788}^{+0.00786}$  &    $-0.00251^{+0.00796}_{-0.00790}$  &    $-0.00232_{-0.00788}^{+0.00783}$  &    $-0.00241 \pm 0.00785$  &    $0.00173_{-0.00750}^{+0.00754}$  \\ 
		\bottomrule
		\bottomrule
	\end{tabular}}
	\caption{\label{table:2} \footnotesize Bounds on cosmological parameters in the NPDDE11+$r$ model. Marginalized limits are given at 68\% C.L. whereas upper limits are given at 95\% C.L.. Note that $H_0$ and $\sigma_8$ are derived parameters.}
\end{table}	

\begin{table}

	\centering 
%	\small\addtolength{\tabcolsep}{-4pt}
    \resizebox{\textwidth}{!}{
	\begin{tabular}{ccccccccc}
		\toprule
		\toprule
		\hspace{1mm}
		 \vspace{0.2cm}
		  Parameter  &    Planck  &    Planck  &    Planck  &    Planck  &    Planck  &    Planck  &    Planck \\ \vspace{0.2cm}
		  &    +BAO  &    +BAO+PAN  &    +BAO+HST+PAN  &    +BAO+HST  &    +HST  &    +HST+PAN  &    +HST+lensing  \\ 
		\toprule
		 \vspace{ 0.2cm}
		  $\Omega_b h^2$   &     $0.02230\pm0.00014$  &    $0.02230\pm0.00014$  &    $0.02237\pm0.00013$  &    $0.02237\pm0.00014$  &    $0.02236\pm0.00015$  &    $0.02237\pm0.00015$  &    $0.02237\pm0.00015$ \\ \vspace{ 0.2cm}
		
		  $\Omega_c h^2$   &     $0.1190\pm0.0010$  &    $0.1188\pm0.0010$  &   $0.1181\pm0.0010$  &    $0.1182\pm0.0010$  &    $0.1183\pm0.0013$  &    $0.1182\pm0.0013$  &    $0.1179\pm 0.0013$ \\ \vspace{ 0.2cm}
		
		  $\tau$   &    $0.083\pm 0.016$  &   $0.084\pm 0.017$  &    $0.088\pm0.016$  &    $0.087\pm0.016$  &    $0.086\pm0.017$  &    $0.089\pm0.017$  &    $0.071\pm0.013$  \\ \vspace{ 0.2cm}
		
		  $n_s$  &    $0.967\pm0.004$  &    $0.967\pm0.004$  &    $0.969\pm0.004$  &    $0.968\pm0.004$  &    $0.968\pm0.004$  &    $0.968\pm0.004$  &    $0.969\pm0.004$  \\ \vspace{ 0.2cm}
		
		  ${\rm{ln}}(10^{10} A_s)$   &    $3.098\pm0.032$  &    $3.100^{+0.033}_{-0.032}$  &    $3.106_{-0.033}^{+0.032}$  &    $3.106\pm0.032$  &    $3.104\pm0.033$  &    $3.107\pm0.033$  &    $3.073\pm0.025$ \\ \vspace{ 0.2cm}
		
		  $\Theta_s$  &    $1.041\pm0.0003$  &    $1.041\pm0.0003$  &    $1.041\pm0.0003$  &    $1.041\pm0.0003$  &    $1.041\pm0.0003$  &    $1.041\pm0.0003$  &    $1.041\pm0.0003$  \\ \vspace{ 0.2cm}
		
		  $H_0$ (km/s/Mpc)  &    $67.63\pm0.47$  &    $67.69\pm0.47$  &   $68.03\pm0.43$  &    $67.99\pm0.45$  &    $67.94_{-0.63}^{+0.62}$  &    $68.01\pm 0.58$  &    $68.13_{-0.61}^{+0.62}$  \\ \vspace{ 0.2cm}
		
		  $\sigma_8$   &   $0.831\pm0.013$  &    $0.831\pm0.013$  &    $0.831\pm0.013$  &    $0.831\pm0.013$  &    $0.831\pm0.013$  &    $0.832\pm0.013$  &    $0.817\pm0.008$  \\ 
		\bottomrule 
		\bottomrule
	\end{tabular}}
	\caption{\label{table:5}\footnotesize Bounds on cosmological parameters in the $\Lambda \textrm{CDM}$ model. Marginalized limits are given at 68\% C.L. whereas upper limits are given at 95\% C.L. Note that $H_0$ and $\sigma_8$ are derived parameters.}
\end{table}

\begin{figure}[tbp]
\centering % \begin{center}/\end{center} takes some additional vertical space
\includegraphics[width=.4963\linewidth]{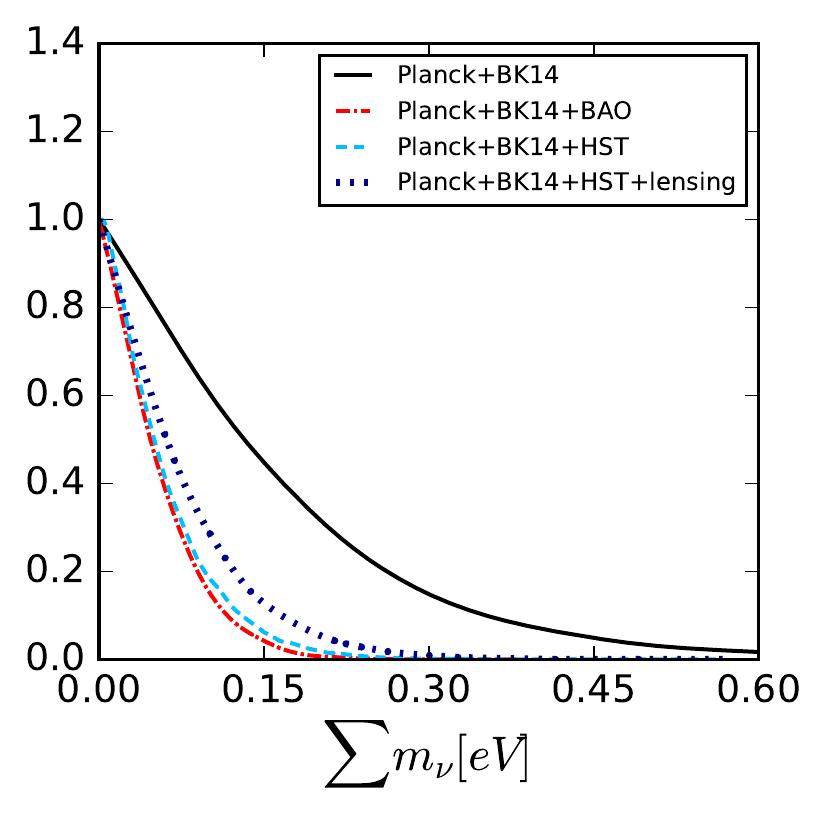}
\hfill
\includegraphics[width=.4963\linewidth]{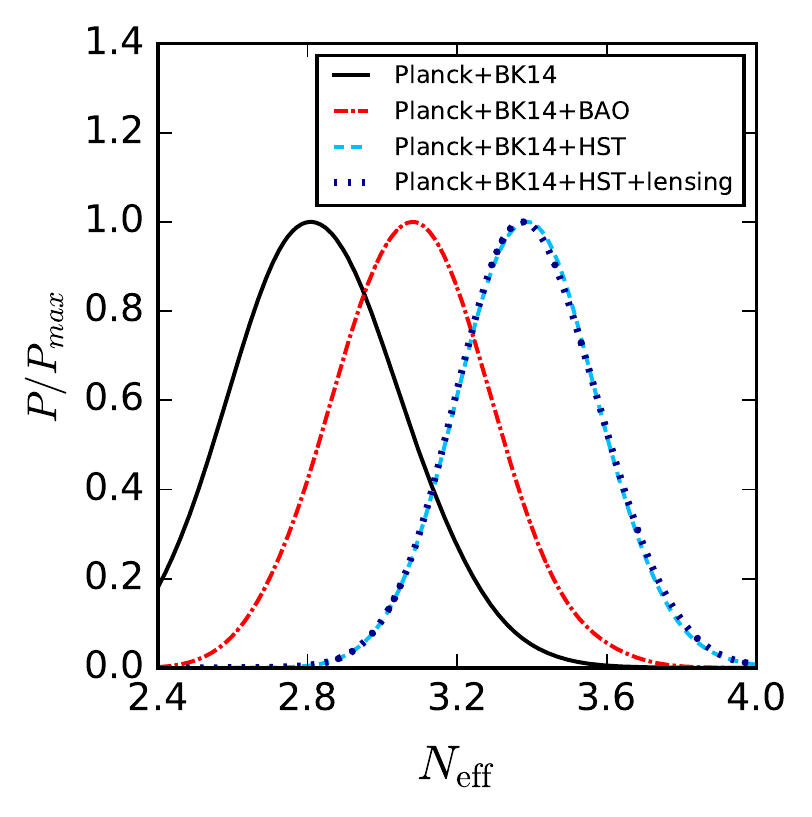}
% "\includegraphics" is very powerful; the graphicx package is already loaded
\caption{\label{fig:1}Comparison of 1-D marginalized posterior distributions for $\sum m_{\nu}$ (eV) and $N_\textrm{eff}$  for various data combinations in NPDDE11+$r$.}
\end{figure}

\begin{figure}[tbp]
\centering % \begin{center}/\end{center} takes some additional vertical space
\includegraphics[width=.4963\linewidth]{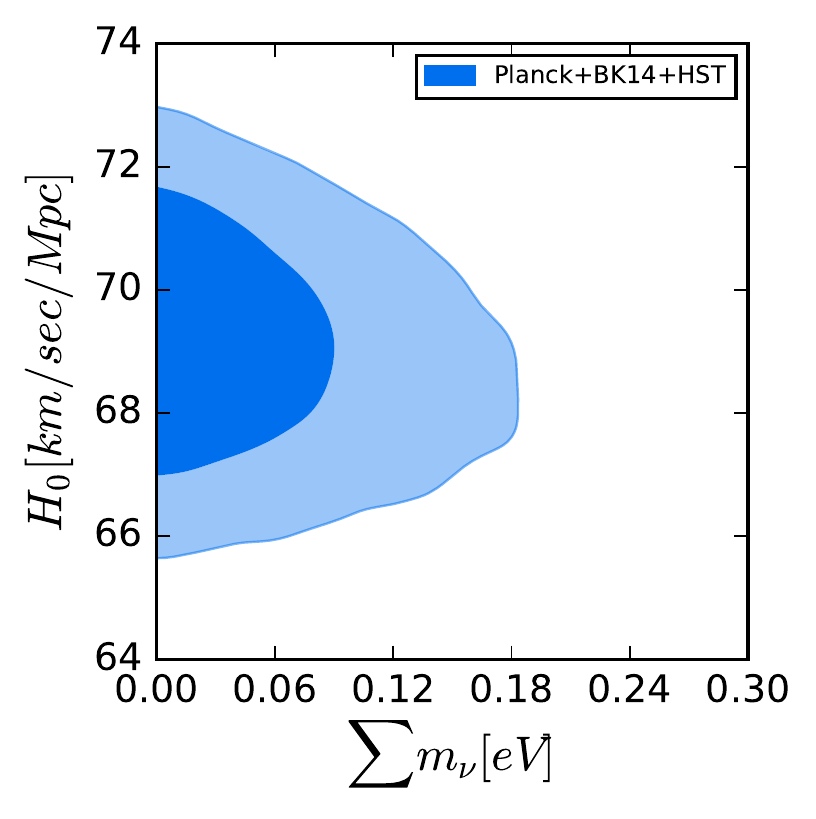}
\hfill
\includegraphics[width=.4963\linewidth]{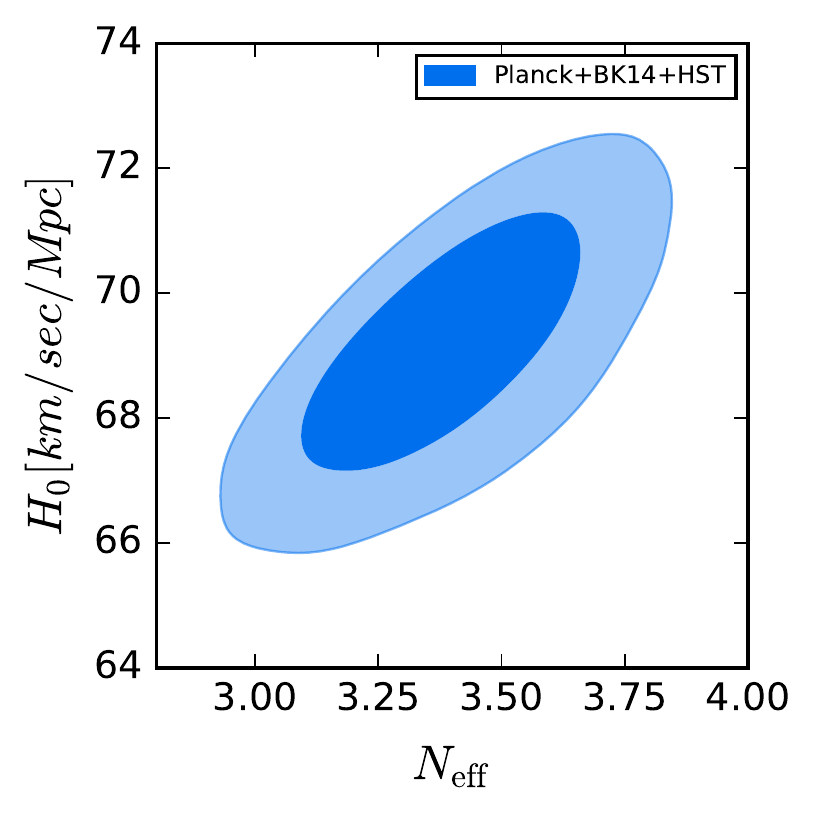}
% "\includegraphics" is very powerful; the graphicx package is already loaded
\caption{\label{fig:2}1$\sigma$ and 2$\sigma$ marginalized  contours for $H_0$ (km/sec/Mpc) vs. $\sum m_{\nu}$ (eV) and $H_0$ (km/sec/Mpc) vs. $N_\textrm{eff}$ for Planck+BK14+HST in the NPDDE11+$r$ model, showing only a small correlation between $H_0$ and $\sum m_{\nu}$ whereas a strong positive correlation between $H_0$ vs. $N_\textrm{eff}$.}
\end{figure}

\begin{figure}[tbp]
\centering % \begin{center}/\end{center} takes some additional vertical space
\includegraphics[width=.5\linewidth]{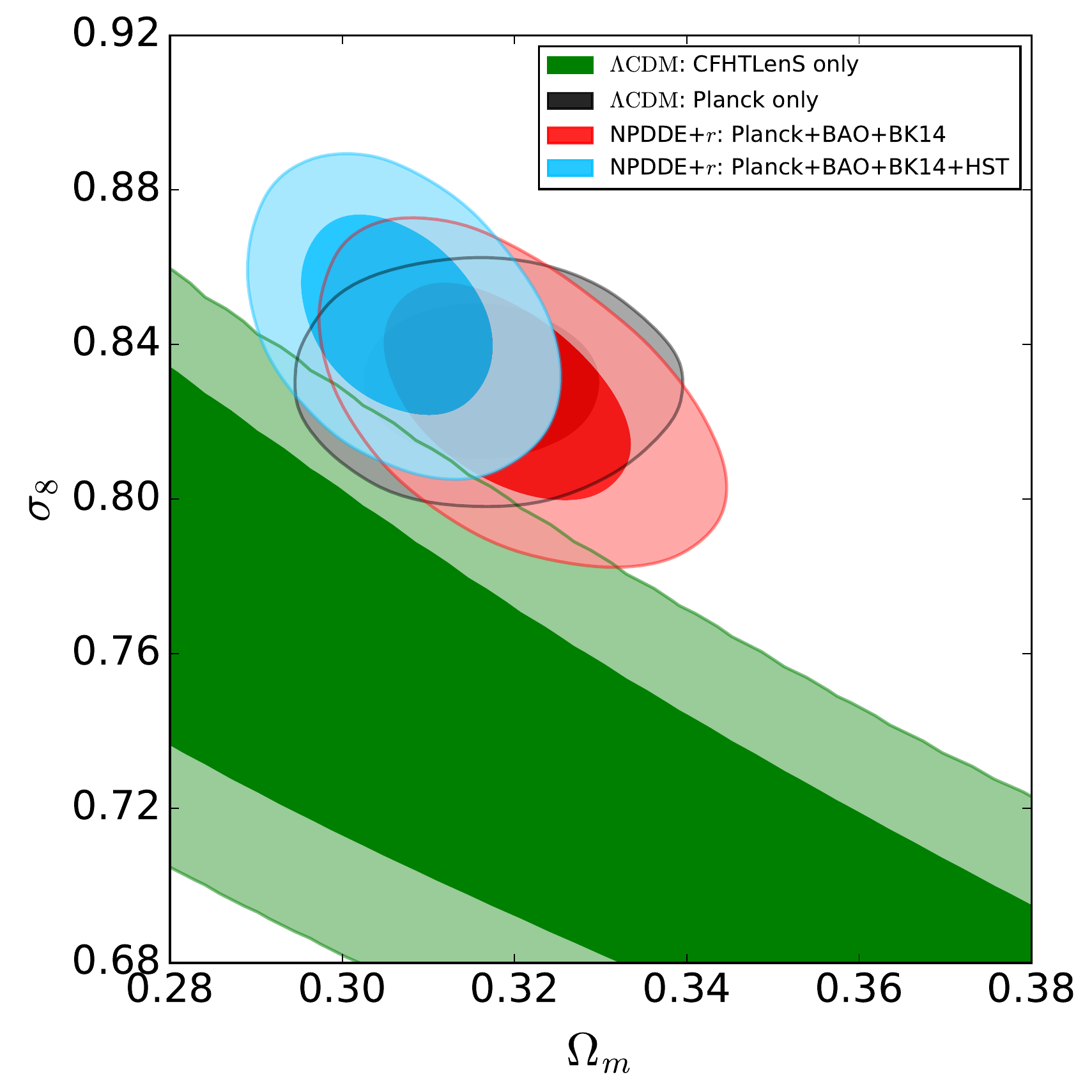}
% "\includegraphics" is very powerful; the graphicx package is already loaded
\caption{\label{fig:7}1$\sigma$ and 2$\sigma$ marginalized  contours in the $\sigma_8-\Omega_m$ plane showing that the NPDDE+$r$ model is ineffective in reducing the tension between CFHTLenS and Planck 2015.}
\end{figure}

\begin{figure}[tbp]
\centering % \begin{center}/\end{center} takes some additional vertical space
\includegraphics[width=.4963\linewidth]{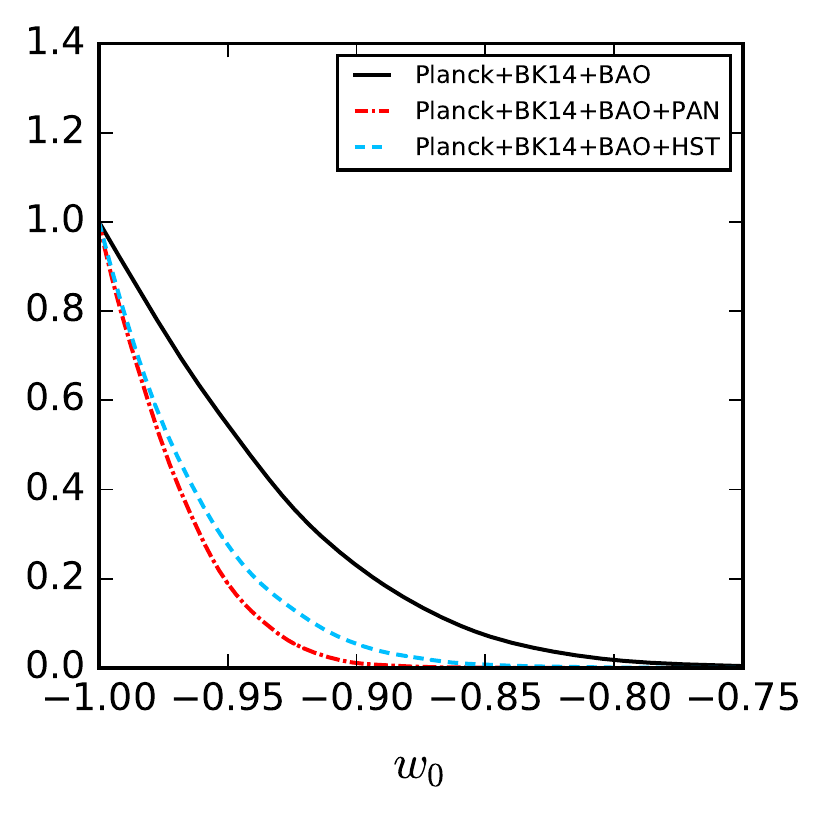}
\hfill
\includegraphics[width=.4963\linewidth]{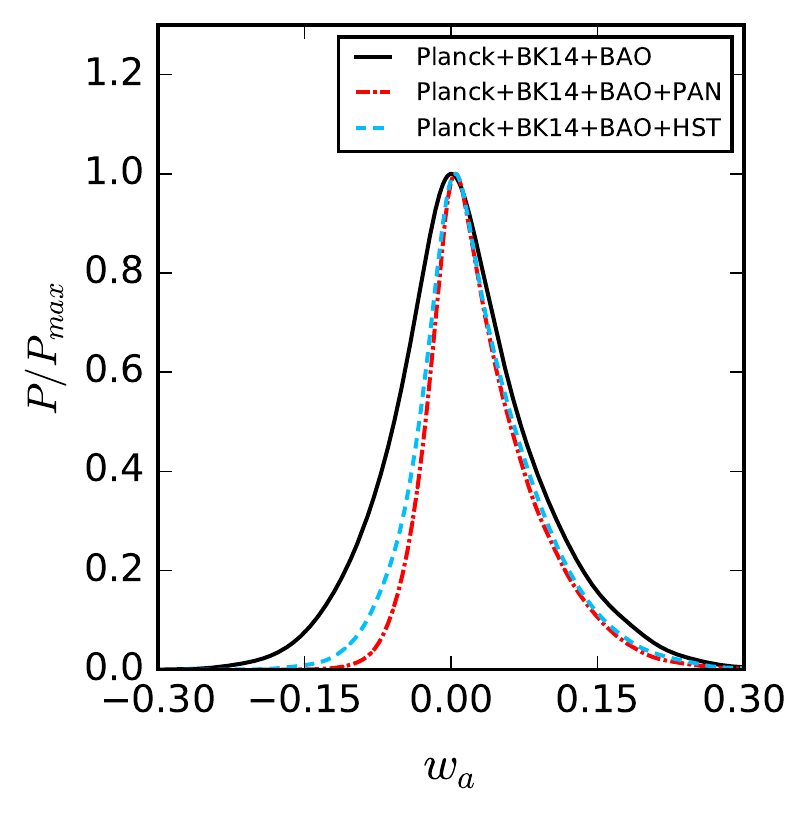}
% "\includegraphics" is very powerful; the graphicx package is already loaded
\caption{\label{fig:3}Comparison of 1-D marginalized posterior distributions for $w_0$ and $w_a$  for different data combinations in NPDDE11+$r$.}
\end{figure}
Bounds on the NPDDE11+$r$ model parameters are presented in table~\ref{table:2} while the bounds on the $\Lambda \textrm{CDM}$ model parameters are presented in table~\ref{table:5}. We do not include the bounds from CMB only data as the bounds are not strong enough in the NPDDE11+$r$ model, a finding that corroborates with a recent study~\cite{DiValentino:2017zyq} which had varied the dark energy EoS in both phantom and non-phantom regions. However adding either BAO or HST with CMB data seems to provide strong bounds on cosmological parameters. Comparing with the bounds on the parameters in the $\Lambda \textrm{CDM}$ model however we can see that the 68\% C.L. spreads of the relevant parameters have increased to different degrees for different parameters. This is an expected phenomenon given the number of parameters has been doubled. Overall the six $\Lambda \textrm{CDM}$ parameters have been estimated in the NPDDE11+$r$ model with reasonable spreads, showing that it is possible to constrain cosmology effectively in a large parameter space with current datasets.

We also find tight bounds on $\sum m_{\nu}$ in this model. The 1-D posteriors for $\sum m_{\nu}$ and $N_\textrm{eff}$ are given in figure~\ref{fig:1}. Our most aggressive bound in this paper is found in this model with Planck+BAO dataset: $\sum m_{\nu}<$ 0.123 eV (95\% C.L.) which is very close to the minimum mass of $\sum m_{\nu} \simeq$ 0.1 eV (95\% C.L.) required for inverted hierarchy of neutrinos (for normal hierarchy, the minimum $\sum m_{\nu}$ required is around 0.06 eV) \cite{1475-7516-2016-11-035}. Although we are in such an extended parameter space, this bound is stronger than a bound of $\sum m_{\nu} <$ 0.158 eV (95\% C.L.)  obtained in $\Lambda \textrm{CDM}+\sum m_{\nu}$ with Planck+BAO \cite{Choudhury:2018byy}. Without the BAO data, only Planck and BK14 together provide a bound of $\sum m_{\nu} < 0.414$ eV (95\% C.L.)  whereas only using Planck in the same model gives us a bound of $\sum m_{\nu} < 0.509$ eV (95\% C.L.)  which is incidentally very close to the bound of $\sum m_{\nu} < 0.49$ eV (95\% C.L.)  reported by Planck collaboration \cite{Ade:2015xua} using the same data in the minimal $\Lambda \textrm{CDM}+\sum m_{\nu}$ model. Recent studies \cite{Vagnozzi:2018jhn,Choudhury:2018byy} in smaller parameter spaces have shown that the models comprising of NPDDE provide stronger bounds on $\sum m_{\nu}$ than $\Lambda \textrm{CDM}+\sum m_{\nu}$, because of a degeneracy present between the dark energy EoS $w$ and $\sum m_{\nu}$ \cite{Hannestad:2005gj} which leads to the phantom region of the dark energy parameter space preferring larger masses and the non-phantom region preferring smaller masses. However, cosmological datasets usually prefer the phantom region more when the dark energy EoS is allowed to vary both in the phantom and non-phantom regions, which usually leads to weaker bounds on $\sum m_{\nu}$. This work shows that even as a 12 parameter model, the NPDDE11+$r$ is very efficient in constraining $\sum m_{\nu}$, unlike the 12 parameter model in \cite{DiValentino:2017zyq}, where the bounds on neutrino mass sum loosens up considerably. Contrary to what happens in lower dimensional parameter spaces, the HST prior does not lead to stronger bounds on $\sum m_{\nu}$, as the magnitude of correlation between $H_0$ and $\sum m_{\nu}$ is very small in this model. 

This small correlation can be explained with the help of mutual degeneracies present between $H_0$, $\sum m_{\nu}$, and the DE EoS $w$. When $w$ is kept constant in a flat $\Lambda CDM+ \sum m_{\nu}$ universe, $H_0$ and $\sum m_{\nu}$ are strongly anti-correlated, to keep the distance to the last scattering surface, $\chi (z_{dec})$ unchanged. Here $z_{dec}$ is the redshift of photon decoupling. $\chi (z_{dec})$ is sensitive to any changes in the values of $H_0$ and $\sum m_{\nu}$, and as shown in \cite{Choudhury:2018byy}, any change to $\chi (z_{dec})$ due to increase in $\sum m_{\nu}$ can be compensated by decreasing $H_0$. This causes the anti-correlation. On the other hand, $H_0$ and $w$ are also degenerate, as both of them control the late time expansion rate of the universe. Thus, when we consider a varying DE EoS, a change in $H_0$ now can be compensated by a change in $w$, instead of $\sum m_{\nu}$. This leads to the decreased degeneracy between $H_0$ and $\sum m_{\nu}$ in our NPDDE models.

However we found a strong positive correlation still present with $N_\textrm{eff}$, which leads to a large increase in the value of $N_{\textrm{eff}}$ with the use of HST prior (the correlations can be visualized in figure~\ref{fig:2}). Indeed, while Planck+BK14+BAO prefers a $H_0 = 66.64^{+1.38}_{-1.37}$ km/sec/Mpc (68\% C.L.), and $N_{\textrm{eff}} = 3.082_{-0.211}^{+0.209}$ (68\% C.L.), the inclusion of the HST prior to this data combination leads to higher values of $H_0 =69.13_{-1.08}^{+1.09}$ km/sec/Mpc (68\% C.L.), and $N_{\textrm{eff}} = 3.392_{-0.186}^{+0.188}$ (68\% C.L.) both. The standard value of $N_{\textrm{eff}} = 3.045$ is excluded at 68\% C.L., and favours a dark radiation component, but only very mildly, since $N_{\textrm{eff}} = 3.045$ is included in 95\% C.L. Thus this exclusion of $N_{\textrm{eff}} = 3.045$ at 68\% C.L. should not be considered as anything of great significance. In this model, this is a general feature in all the dataset combinations that have the HST prior included, solely due the large tension present between Planck and HST. The HST prior also prefers higher values of $\sigma_8$. This model does not help the conflict between Planck and CFHTLenS regarding the value of $\sigma_8$. Visual depiction of this can be found in figure~\ref{fig:7} in the $\sigma_8-\Omega_m$ plane. Inclusion of the lensing data lead to worsening of the mass bounds whereas bounds on $N_\textrm{eff}$ are almost unaffected. These datasets however lower the preferred $\sigma_8$ values.\\ 

The use of the $H071p6$ prior, which has a lower value of $H_0$ than HST, however, leads to lower values
of $N_{\textrm{eff}}$, due to a smaller tension between Planck and $H071p6$. In particular, with Planck + BK14 + BAO + $H07106$, we get a bound of $N_{\textrm{eff}} = 3.202^{+0.200}_{-0.202}$ (68\%). Thus, $N_{\textrm{eff}} = 3.045$ is no longer 
excluded at 68\% in this case. \\

The SNe Ia luminosity distance measurements provide information about  evolution of luminosity distance as a function of redshift ($0.01<z<2.3$ for the Pantheon sample). This can be used to measure the evolution of the scale factor \cite{Astier:2000as} and is helpful in constraining the dark energy EoS. We found that addition of the PAN data did help in constraining the dark energy parameters more tightly. For Planck+BK14+BAO, we have a bound of $w_0 < -0.859$ (95\% C.L.), which shrinks to $w_0 < -0.933$ (95\% C.L.) with the addition of PAN. On the other hand, Planck+BK14+BAO produces a bound of $w_a = 0.013^{+0.065}_{-0.077}$ (68\% C.L.), whereas Planck+BK14+BAO+PAN leads to $w_a = 0.033^{+0.036}_{-0.063}$ (68\% C.L.). We see that the 68\% spreads of $w_a$ have shrunk. This has also been depicted in figure~\ref{fig:3}. The HST prior also has similar but less strong effect. With Planck+BK14+BAO+HST we have $w_0 < -$0.908 (95\% C.L.) and $w_a = 0.028^{+0.046}_{-0.065}$ (68\% C.L.). In all cases we found that the cosmology is compatible with a cosmological constant (i.e., $w_0 = -1$, $w_a = 0$). 

As far as values of the tensor-to-scalar ratio is concerned, we find that if we run the chains without the BK14 data, we get a bound of $r_{0.05} <$ 0.155 (95\% C.L.) with Planck+BAO, which is higher than the bound of $r_{0.05} <$ 0.12 (95\% C.L.) set by Planck collaboration \cite{Ade:2015xua}. However, inclusion of the BK14 data leads to a bound of $r_{0.05} <$ 0.075 (95\% C.L.), which is close to the $r_{0.05} < 0.07$ (95\% C.L.) limit set by the BICEP2/Keck collaboration \cite{Array:2015xqh}. The value of $r_{0.05}$ remains almost unchanged across all the datasets as long as the BK14 data is included.

\subsection{NPDDE11 model}
\label{sec:4:2}

\begin{table}

	\centering 
%	\small\addtolength{\tabcolsep}{-4pt}
    \resizebox{\textwidth}{!}{
	\begin{tabular}{ccccccccc}
		\toprule
		\toprule
		\hspace{1mm}
		 \vspace{0.2cm}
		  Parameter  &    Planck  &    Planck  &    Planck  &    Planck  &    Planck  &    Planck  &    Planck \\ \vspace{0.2cm}
		  &    +BAO  &    +BAO+PAN  &    +BAO+HST+PAN  &    +BAO+HST  &    +HST  &    +HST+PAN  &    +HST+lensing  \\ 
		\toprule
		 \vspace{ 0.2cm}
		  $\Omega_b h^2$   &    $0.02241\pm0.00021$  &    $0.02242\pm0.00020$  &    $0.02264\pm0.00018$  &    $0.02264\pm0.00019$  &    $0.02264\pm0.00020$  &    $0.02266\pm0.00020$  &    $0.02261\pm0.00020$  \\ \vspace{ 0.2cm}
		
		  $\Omega_c h^2$   &    $0.1187\pm0.0033$  &    $0.1188\pm0.0034$  &    $0.1230\pm0.0031$  &    $0.1232\pm0.0031$  &    $0.1232\pm0.0031$  &    $0.1230_{-0.0030}^{+0.0031}$  &    $0.1226\pm 0.0031$ \\ \vspace{ 0.2cm}
		
		  $\tau$   &    $0.092\pm0.018$  &    $0.091\pm0.018$  &    $0.095\pm 0.018$  &    $0.095\pm0.018$  &    $0.095\pm0.018$  &    $0.096\pm0.018$  &    $0.077^{+0.014}_{-0.016}$ \\ \vspace{ 0.2cm}
		
		  $n_s$  &    $0.969\pm0.009$  &    $0.969\pm0.009$  &    $0.981\pm0.008$  &    $0.981\pm0.008$  &    $0.981\pm0.009$  &    $0.981\pm0.008$  &    $0.982\pm0.009$      \\ \vspace{ 0.2cm}
		
		  ${\rm{ln}}(10^{10} A_s)$   &    $3.117^{+0.037}_{-0.038}$  &    $3.116\pm0.038$  &    $3.134^{+0.036}_{-0.037}$  &    $3.134_{-0.038}^{+0.037}$  &    $3.133\pm0.037$  &    $3.135_{-0.037}^{+0.036}$   \\ \vspace{ 0.2cm}
		
		  $\Theta_s$  &    $1.041\pm0.0005$  &    $1.041\pm0.0005$  &    $1.040\pm0.0004$  &    $1.040\pm0.0004$  &    $1.040\pm0.0004$  &    $1.040\pm0.0004$  &    $1.041\pm0.0004$  \\ \vspace{ 0.2cm}
		
		  $H_0$ (km/s/Mpc)  &    $66.53_{-1.36}^{+1.37}$  &    $67.32_{-1.28}^{+1.27}$  &    $69.35^{+1.06}_{-1.04}$  &    $69.09_{-1.11}^{+1.10}$  &    $69.07_{-1.38}^{+1.39}$  &    $69.56\pm 1.22$  &    $69.05\pm1.39$  \\ \vspace{ 0.2cm}
		
		  $\sigma_8$   &    $0.822\pm0.019$  &    $0.829\pm0.018$  &    $0.846\pm0.017$  &    $0.844\pm0.018$  &    $0.843\pm0.019$  &    $0.847\pm0.018$  &    $0.823_{-0.014}^{+0.015}$  \\ \vspace{ 0.2cm}
		
		  $\sum m_\nu$ (eV)  &   $<0.126$ &    $<0.128$  &    $<0.137$  &    $<0.131$  &    $<0.151$  &    $<0.135$  &    $<0.191$  \\ \vspace{ 0.2cm}
		
		  $w_0$  &    $<-0.851$  &    $<-0.934$  &    $<-0.941$  &    $<-0.909$  &    $<-0.912$  &    $<-0.944$  &    $<-0.914$  \\ \vspace{ 0.2cm}
		
		  $w_a$  &    $0.011^{+0.069}_{-0.079}$  &    $0.035^{+0.035}_{-0.064}$  &    $0.035_{-0.060}^{+0.032}$   &    $0.030_{-0.066}^{+0.046}$  &    $0.030_{-0.064}^{+0.043}$  &    $0.032_{-0.055}^{+0.029}$  &    $0.035_{-0.069}^{+0.042}$  \\ \vspace{ 0.2cm}
		
		  $N_{\textrm{eff}}$  &    $3.073^{+0.209}_{-0.211}$  &    $3.081_{-0.211}^{+0.212}$  &    $3.378_{-0.184}^{+0.185}$  &    $3.389_{-0.194}^{+0.178}$  &    $3.385\pm 0.190$  &    $3.388_{-0.0182}^{+0.0183}$  &    $3.382\pm0.191$  \\ \vspace{ 0.2cm}
		
		  $n_{\rm run}$   &    $-0.00511^{+0.00775}_{-0.00780}$  &    $-0.00477_{-0.00784}^{+0.00785}$  &    $-0.00027_{-0.00768}^{+0.00770}$   &    $-0.00012^{+0.00777}_{-0.00783}$&    $-0.00016^{+0.00775}_{-0.00777}$  &    $-0.00003_{-0.00776}^{+0.00783}$  &    $0.00356\pm0.00742$  \\ 
		\bottomrule 
		\bottomrule
	\end{tabular}}
	\caption{\label{table:3}\footnotesize Bounds on cosmological parameters in the NPDDE11 model. Marginalized limits are given at 68\% C.L. whereas upper limits are given at 95\% C.L.. Note that $H_0$ and $\sigma_8$ are derived parameters.}
\end{table}

\begin{figure}[tbp]
\centering % \begin{center}/\end{center} takes some additional vertical space
\includegraphics[width=.4963\linewidth]{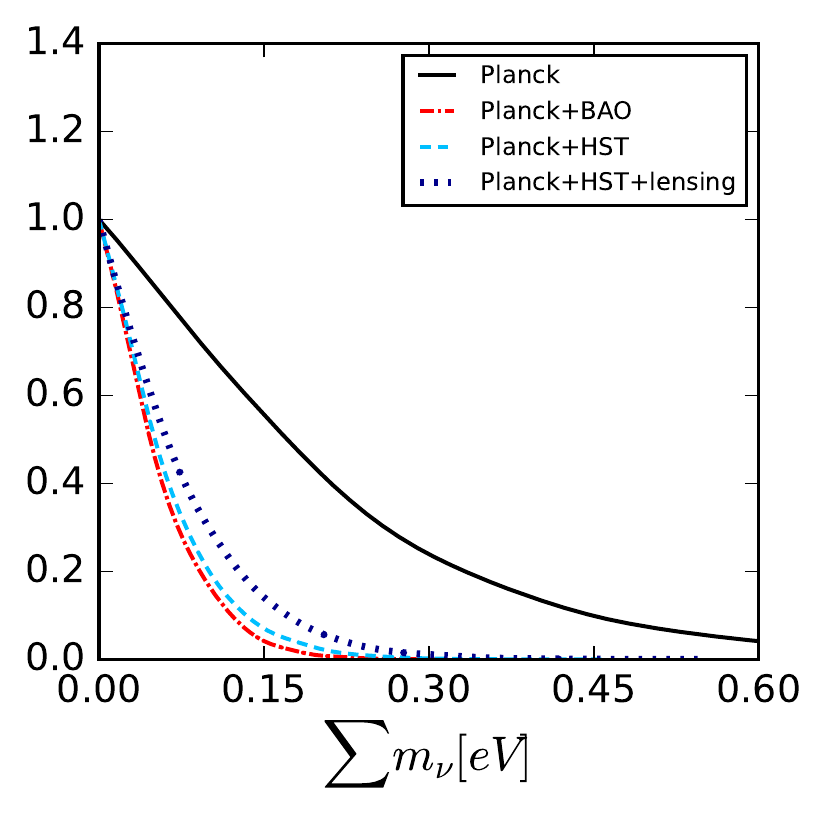}
\hfill
\includegraphics[width=.4963\linewidth]{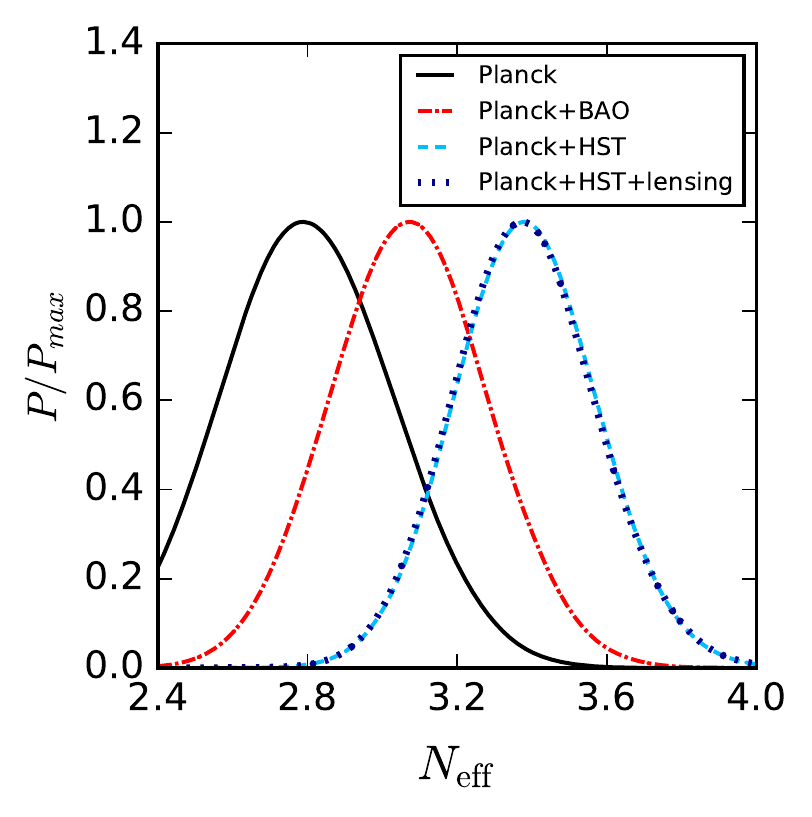}
% "\includegraphics" is very powerful; the graphicx package is already loaded
\caption{\label{fig:4}Comparison of 1-D marginalized posterior distributions for $\sum m_{\nu}$ (eV) and $N_\textrm{eff}$  for various data combinations in NPDDE11.}
\end{figure}

\begin{figure}[tbp]
\centering % \begin{center}/\end{center} takes some additional vertical space
\includegraphics[width=.4963\linewidth]{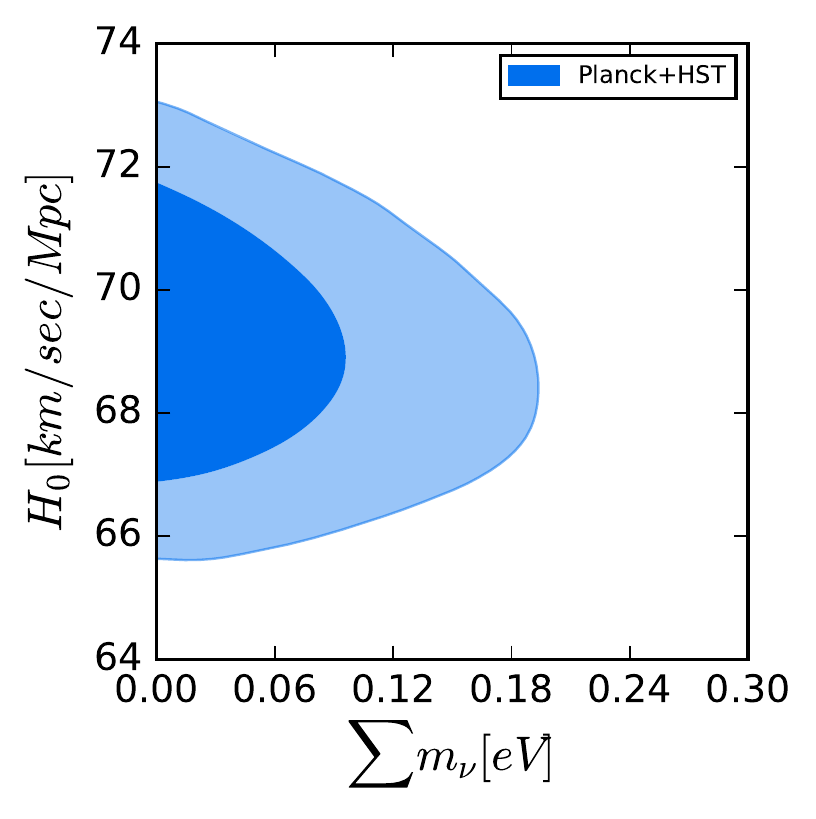}
\hfill
\includegraphics[width=.4963\linewidth]{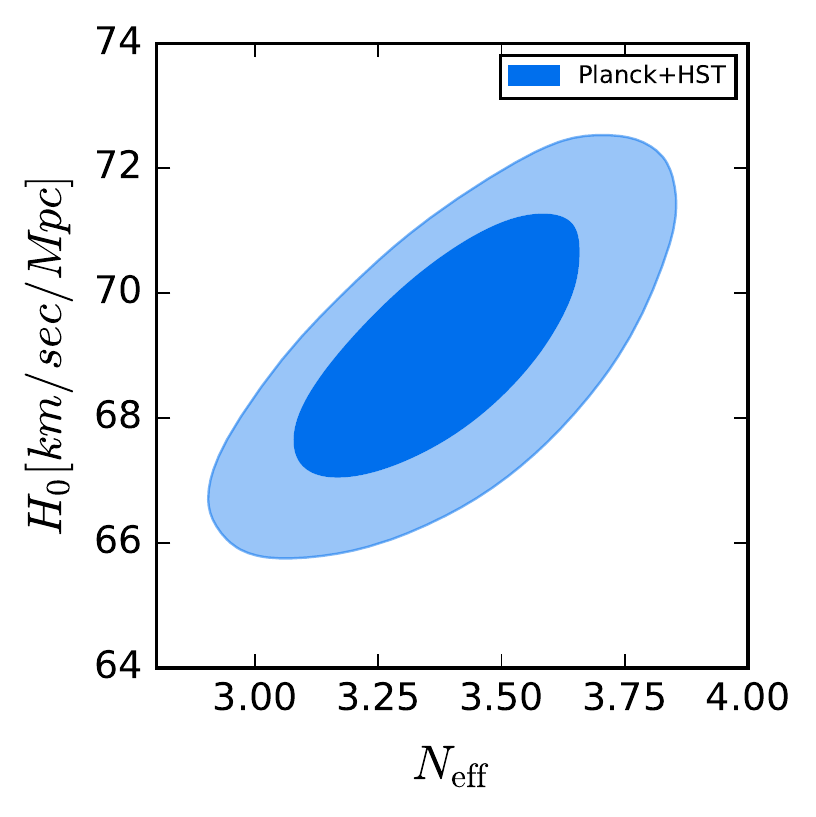}
% "\includegraphics" is very powerful; the graphicx package is already loaded
\caption{\label{fig:5}1$\sigma$ and 2$\sigma$ marginalized  contours for $H_0$ (km/sec/Mpc) vs. $\sum m_{\nu}$ (eV) and $H_0$ (km/sec/Mpc) vs. $N_\textrm{eff}$ for Planck+HST in the NPDDE11 model, showing negligible correlation between $H_0$ and $\sum m_{\nu}$ whereas a strong positive correlation between $H_0$ vs. $N_\textrm{eff}$.}
\end{figure}
\begin{figure}[tbp]
\centering % \begin{center}/\end{center} takes some additional vertical space
\includegraphics[width=.4963\linewidth]{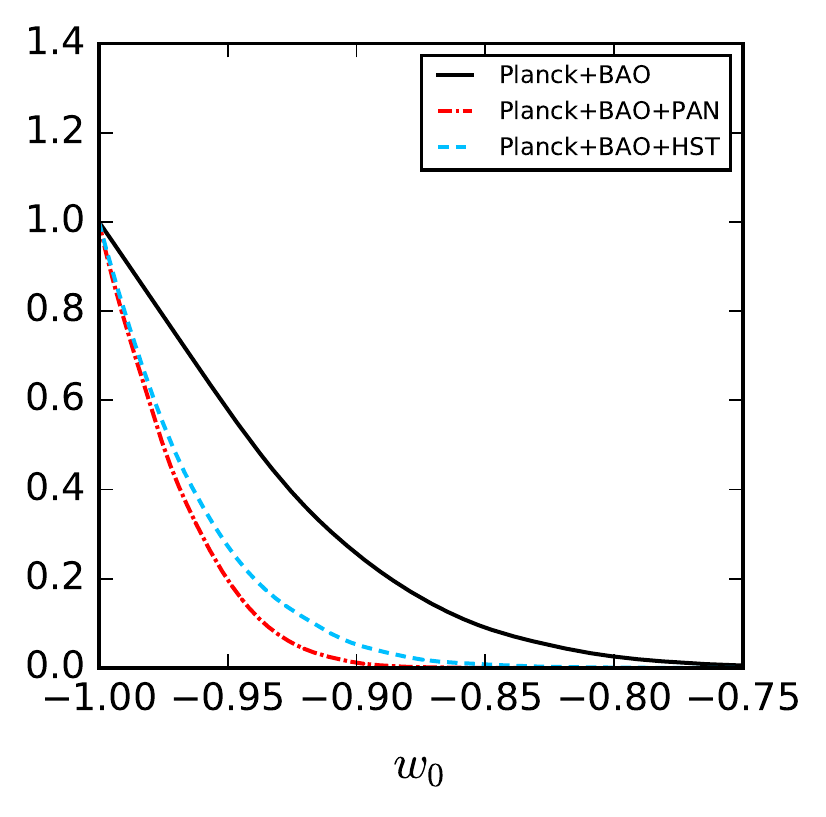}
\hfill
\includegraphics[width=.4963\linewidth]{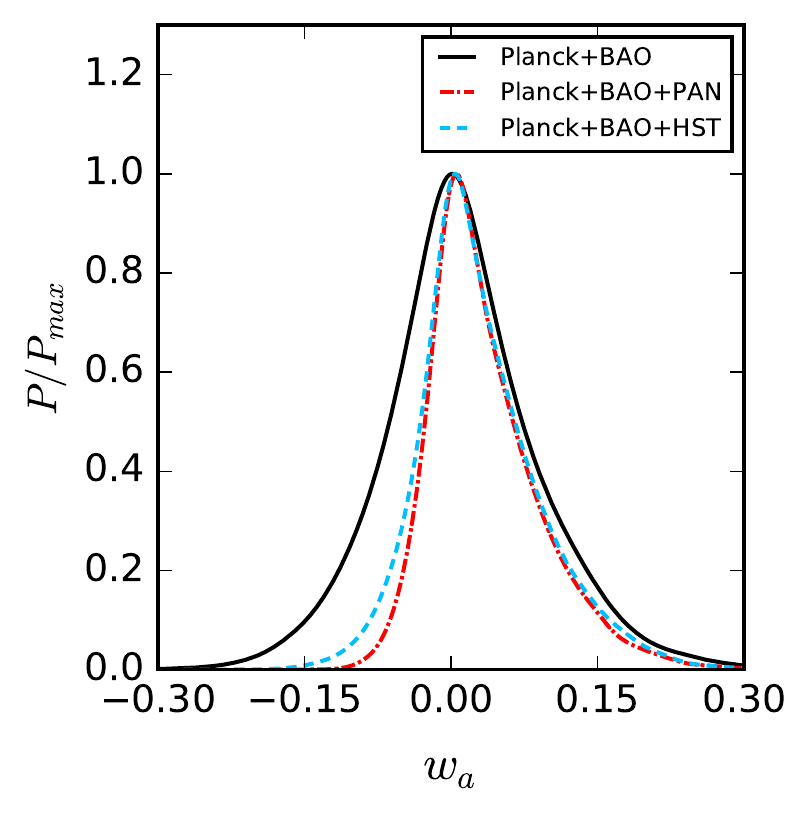}
% "\includegraphics" is very powerful; the graphicx package is already loaded
\caption{\label{fig:6}Comparison of 1-D marginalized posterior distributions for $w_0$ and $w_a$  for different data combinations in NPDDE11.}
\end{figure}
In this section we consider the NPDDE11 model where we turn off the tensor perturbations and also do not include the BK14 data. This does not affect the bounds much as can be seen from table~\ref{table:3} and comparing with table~\ref{table:2}, which verifies the stability of the results in a smaller parameter space. 

The 1-D posteriors for $\sum m_{\nu}$ and $N_{\textrm{eff}}$ for selected datasets are given in figure~\ref{fig:4}. We again find strong bounds on the sum of neutrino masses. We notice that the removal of BK14 data has a small effect on $\sum m_{\nu}$ which persists over different datasets. For instance, in NPDDE11+$r$, for Planck+BAO, we find a $\sum m_{\nu} < 0.131$ eV (95\% C.L.), which is reduced to $\sum m_{\nu} < 0.123$ eV (95\% C.L.)  when we add the BK14 data. In the NPDDE11, this bound is $\sum m_{\nu} < 0.126$ eV (95\% C.L.)  with Planck+BAO, which is our best bound in this model. This is also stronger than the bound obtained in $\Lambda \textrm{CDM}+\sum m_{\nu}$ with Planck+BAO, as in the previous NPDDE11+$r$ model, and a large improvement compared to the ones presented in \cite{DiValentino:2017zyq}, which varied dark energy parameters in both in phantom and non-phantom range. 

The strengthening of the bound from NPDDE11+$r$ to NPDDE11 with Planck+BAO might simply be due to reduction in the parameter space volume. On the other hand it seems BK14 prefers a lower $\sum m_{\nu}$. However even then the changes are small. BK14 data also seems to prefer slightly larger values of $\sigma_8$, thereby increasing the tension with CFHTLenS. Also, the inclusion of HST prior again seems to discard the standard value of $N_{\textrm{eff}} = 3.045$ at 68\% C.L. but again, not at 95\% C.L., and also it doesn't lead to stronger $\sum m_{\nu}$, as before in the NPDDE+$r$ model, due to a large positive correlation between $H_0$ and $N_{\textrm{eff}}$ but a only small correlation between $H_0$ and $\sum m_{\nu}$. This can be visualized in figure~\ref{fig:5}. The PAN dataset provides stricter bounds on $w_0$ and $w_a$, as before. We depict that in figure~\ref{fig:6}.\\

The use of the $H071p6$ prior instead of HST, here again, leads to lower values
of $N_{\textrm{eff}}$. For instance, with Planck+BAO+$H07106$, we get a bound of $N_{\textrm{eff}} = 3.193^{+0.197}_{-0.199}$ (68\%). Thus, $N_{\textrm{eff}} = 3.045$ is no longer  excluded at 68\% in this model also.\\ 

\subsection{NPDDE11+$A_{\textrm{lens}}$ model}

\label{sec:4:3}
\begin{table}

	\centering
%	\small\addtolength{\tabcolsep}{-4pt}
    \resizebox{\textwidth}{!}{
	\begin{tabular}{ccccccccc}
		\toprule
		\toprule
		 \vspace{0.2cm}
		  Parameter  &    Planck  &    Planck  &    Planck  &    Planck  &    Planck  &    Planck  &    Planck  \\\vspace{0.2cm}
		  &    +BAO  &   +BAO+PAN  &    +BAO+HST+PAN  &    +BAO+HST   &    +HST  &    +HST+PAN  &    +HST+lensing  \\
		\midrule
		\hspace{1mm}
		\vspace{ 0.2cm}
		  $\Omega_b h^2$  &    $0.02265\pm0.00024$  &    $0.02263\pm0.00023$  &    $0.02285\pm0.00021$  &    $0.02288^{+0.00021}_{-0.00024}$  &    $0.02289\pm0.00023$  &    $0.02289\pm0.00022$  &    $0.02270\pm0.00021$  \\ \vspace{ 0.2cm}
		
		  $\Omega_c h^2$  &    $0.1192\pm0.0034$  &    $0.1192^{+0.0034}_{-0.0033}$  &    $0.1231\pm0.0031$  &    $0.1234_{-0.0031}^{+0.0032}$  &    $0.1232\pm0.0032$  &    $0.1228\pm0.0031$  &    $0.1226_{-0.0033}^{+0.0030}$  \\ \vspace{ 0.2cm}
		
	 	$\tau$  &    $0.059^{+0.021}_{-0.022}$  &    $0.059^{+0.021}_{-0.022}$  &    $0.059^{+0.021}_{-0.022}$  &    $0.059 \pm 0.022$  &      $0.059 \pm 0.022$  &    $0.060 \pm 0.022$  &    $0.058^{+0.021}_{-0.022}$  \\ \vspace{ 0.2cm}
		
	 	$n_s$  &    $0.978\pm0.011$  &    $0.978\pm0.010$  &    $0.989\pm0.009$  &    $0.991\pm 0.009$  &    $0.991\pm0.010$  &    $0.991\pm 0.009$  &    $0.986\pm0.009$  \\ \vspace{ 0.2cm}
		
	 	${\rm{ln}}(10^{10} A_s)$  &    $3.052\pm0.044$  &    $3.052^{+0.044}_{-0.045}$  &    $3.060^{+0.044}_{-0.045}$  &    $3.060_{-0.045}^{+0.044}$  &     $3.060^{+0.044}_{-0.045}$  &    $3.060_{-0.045}^{+0.044}$  &    $3.055_{-0.044}^{+0.043}$  \\ \vspace{ 0.2cm}
		
	 	$\Theta_s$  &    $1.041\pm0.0005$  &    $1.041\pm0.0005$  &    $1.040\pm0.0004$  &    $1.040\pm0.0004$  &    $1.040\pm0.0004$  &    $1.040\pm0.0004$  &    $1.041\pm0.0004$  \\ \vspace{ 0.2cm}
		
	 	$H_0$ (km/s/Mpc)  &    $66.99^{+1.45}_{-1.46}$  &    $67.94\pm1.30$  &    $69.84^{+1.07}_{-1.07}$  &     $69.52\pm 1.13$   &    $69.80\pm1.48$  &    $70.32_{-1.29}^{+1.28}$  &    $69.23_{-1.45}^{+1.44}$  \\ \vspace{ 0.2cm}
		
	 	$\sigma_8$  &    $0.781\pm0.025$  &    $0.791^{+0.025}_{-0.023}$  &    $0.799^{+0.027}_{-0.024}$  &    $0.795^{+0.027}_{-0.024}$  &    $0.796_{-0.024}^{+0.030}$  &    $0.802_{-0.023}^{+0.027}$  &    $0.795_{-0.023}^{+0.030}$  \\ \vspace{ 0.2cm}
		
	 	$\sum m_\nu$ (eV)  &    $<0.239$  &    $<0.246$  &    $<0.278$  &    $<0.272$  &    $<0.312$  &    $<0.269$  &    $<0.321$ \\ \vspace{ 0.2cm}
		
	 	$w_0$  &    $<-0.812$  &    $<-0.923$  &    $<-0.930$  &    $<-0.875$  &    $<-0.890$  &    $<-0.933$  &    $<-0.903$ \\ \vspace{ 0.2cm}
		
	 	$w_a$  &    $0.020^{+0.089}_{-0.114}$  &    $0.056^{+0.048}_{-0.089}$  &    $0.057_{-0.086}^{+0.044}$  &    $0.052_{-0.102}^{+0.069}$  &    $0.048_{-0.092}^{+0.056}$  &    $0.047_{-0.077}^{+0.038}$  &    $0.043_{-0.083}^{+0.047}$  \\ \vspace{ 0.2cm}
		
	 	$N_{\textrm{eff}}$  &    $3.212^{+0.227}_{-0.228}$  &    $3.201\pm 0.223$  &    $3.487_{-0.210}^{+0.192}$  &    $3.519_{-0.222}^{+0.195}$  &    $3.517_{-0.216}^{+0.196}$  &    $3.497\pm0.197$  &    $3.440_{-0.210}^{+0.192}$  \\ \vspace{ 0.2cm}
		
	 	$n_{\rm run}$  &    $0.00136^{+0.00806}_{-0.00807}$  &    $0.00123^{+0.00805}_{-0.00809}$  &    $0.00638_{-0.00778}^{+0.00782}$  &    $0.00676^{+0.00791}_{-0.00790}$  &    $0.00685_{-0.00803}^{+0.00794}$  &    $0.00643\pm0.00778$  &    $0.00718\pm0.00788$  \\ \vspace{ 0.2cm}
		
	 	$A_{\textrm{lens}}$  &    $1.21^{+0.08}_{-0.09}$  &    $1.20^{+0.08}_{-0.09}$  &    $1.23_{-0.10}^{+0.08}$  &    $1.24_{-0.10}^{+0.08}$  &    $1.24_{-0.10}^{+0.08}$  &    $1.24_{-0.10}^{+0.08}$  &    $1.08_{-0.07}^{+0.06}$  \\
		
		\bottomrule
		\bottomrule
	\end{tabular}}
	\caption{\label{table:4}\footnotesize Bounds on cosmological parameters in the NPDDE11+$A_{\textrm{lens}}$ model. Marginalized limits are given at 68\% C.L. whereas upper limits are given at 95\% C.L. Note that $H_0$ and $\sigma_8$ are derived parameters.}
\end{table}	

\begin{figure}[tbp]
\centering % \begin{center}/\end{center} takes some additional vertical space
\includegraphics[width=.4963\linewidth]{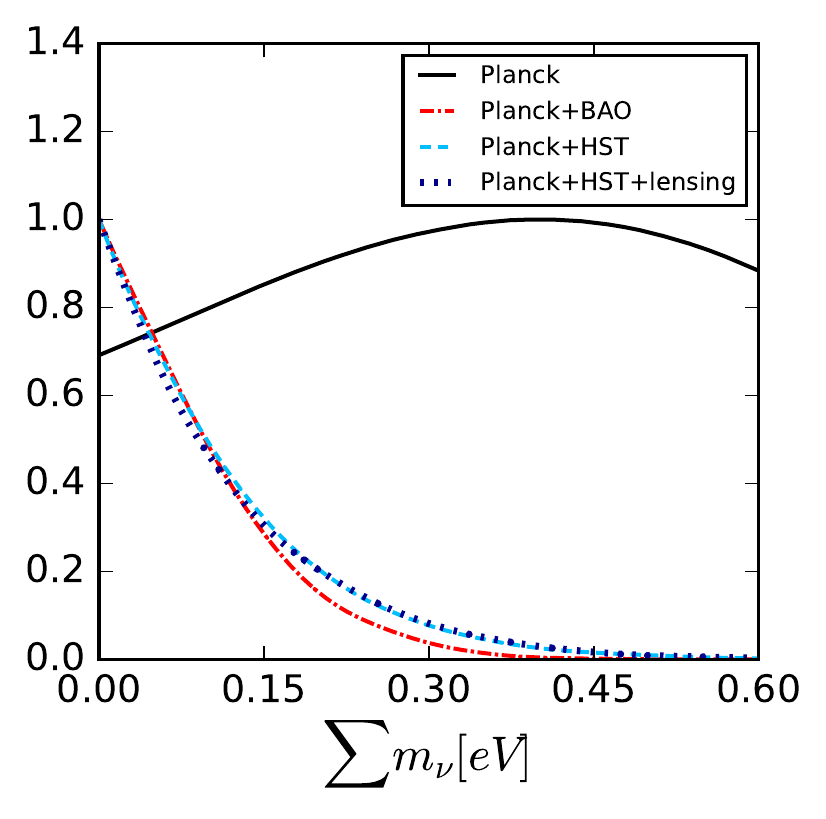}
\hfill
\includegraphics[width=.4963\linewidth]{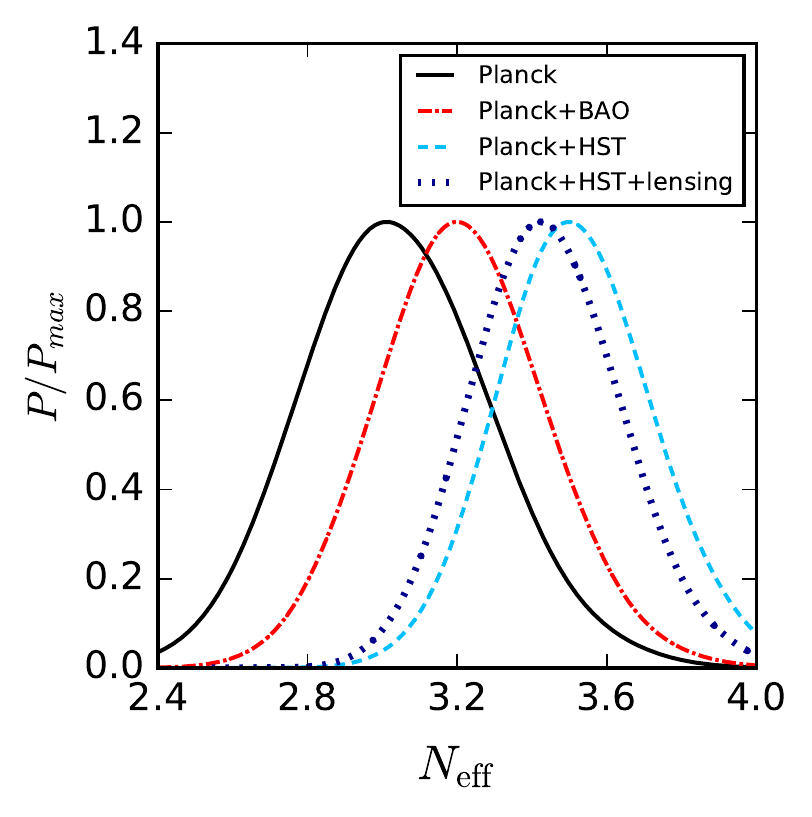}
% "\includegraphics" is very powerful; the graphicx package is already loaded
\caption{\label{fig:9}Comparison of 1-D marginalized posterior distributions for $\sum m_{\nu}$ (eV) and $N_\textrm{eff}$  for various data combinations in NPDDE11+$A_{\textrm{lens}}$.}
\end{figure}

%\begin{figure}[tbp]
%\centering % \begin{center}/\end{center} takes some additional vertical space
%\includegraphics[width=.4963\linewidth]{h0_mnu_planck_r16_npdde11+alens.pdf}
%\hfill
%\includegraphics[width=.4963\linewidth]{h0_nnu_planck_r16_npdde11+alens.pdf}
%% "\includegraphics" is very powerful; the graphicx package is already loaded
%\caption{\label{fig:10}1$\sigma$ and 2$\sigma$ marginalised  contours for $H_0$ (km/sec/Mpc) vs. $\sum m_{\nu}$ (eV) and $H_0$ (km/sec/Mpc) vs. $N_\textrm{eff}$ for Planck+HST in the NPDDE11+$A_{\textrm{lens}}$ model, showing negligible correlation between $H_0$ and $\sum m_{\nu}$ whereas a strong positive correlation between $H_0$ vs. $N_\textrm{eff}$.}
%\end{figure}

\begin{figure}[tbp]
\centering % \begin{center}/\end{center} takes some additional vertical space
\includegraphics[width=.5\linewidth]{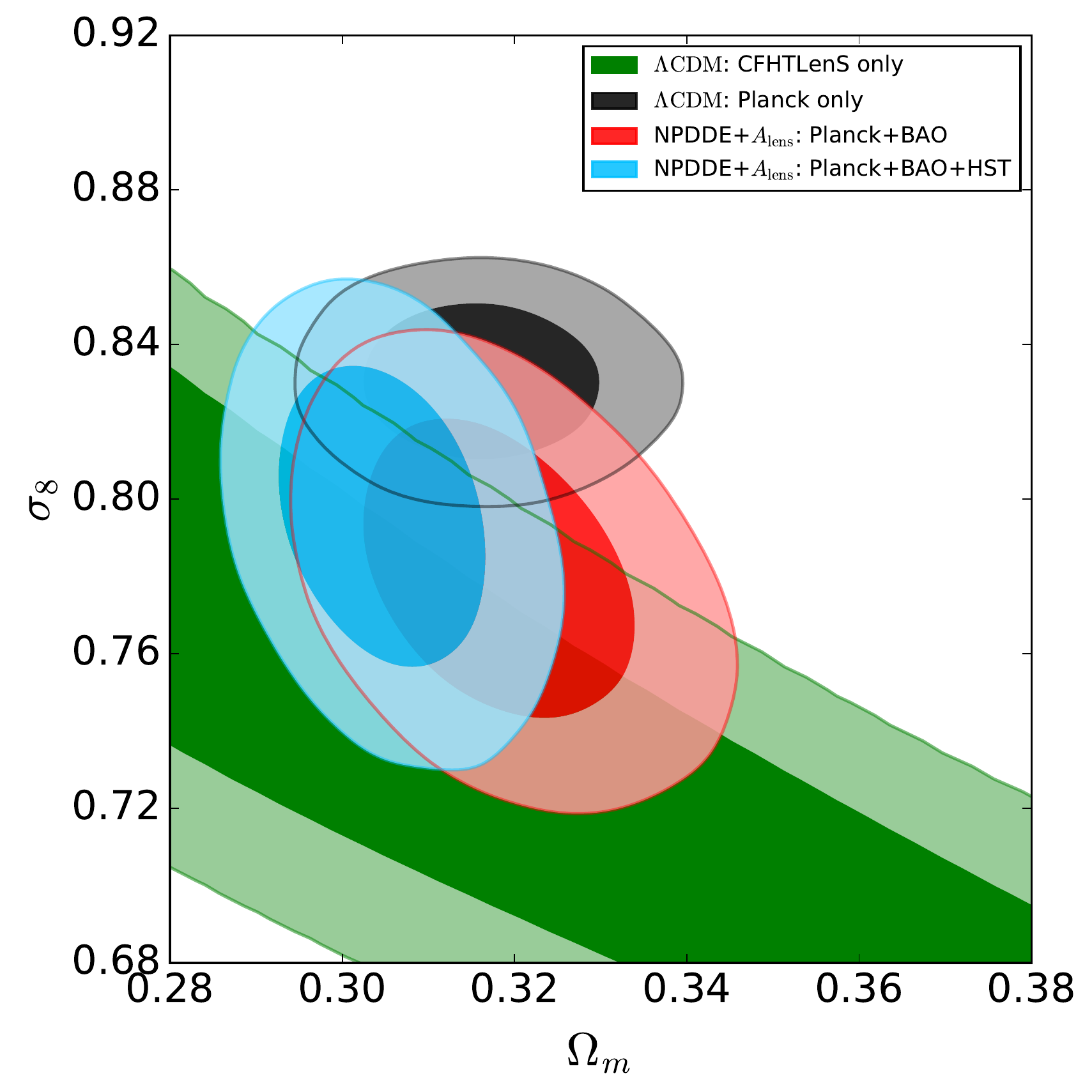}
% "\includegraphics" is very powerful; the graphicx package is already loaded
\caption{\label{fig:8}1$\sigma$ and 2$\sigma$ marginalized  contours in the $\sigma_8-\Omega_m$ plane showing that the NPDDE11+$A_{\textrm{lens}}$ model is effective in reducing the tension between CFHTLenS and Planck 2015.}
\end{figure}

We present the limits on the cosmological parameters in table~\ref{table:4}. A number of important changes happen with the introduction of the new varying parameter $A_{\textrm{lens}}$. Considering that our main goal in this paper is to constrain neutrino masses, we see a substantial relaxation in the bounds on $\sum m_{\nu}$. In previous cases we had fixed $A_{\textrm{lens}} =$ 1. However now that $A_{\textrm{lens}}$ is varied we find that the data prefers a large $A_{\textrm{lens}}$ and discards the $\Lambda \textrm{CDM}$ value of $A_{\textrm{lens}} = 1$ at more than 95\% C.L. (except in case of inclusion of Planck lensing data, which prefers a much lower $A_{\textrm{lens}}$, implying a tension between Planck and lensing). The increasing of the lensing amplitude $A_{\textrm{lens}}$ has the same effect as the decreasing of $\sum m_{\nu}$ \cite{Allison:2015qca}. Increasing $A_{\textrm{lens}}$ leads to smearing of high-$l$ peaks in the CMB temperature and polarization angular power spectra ($C_l^{TT}$, $C_l^{TE}$, $C_l^{EE}$, $C_l^{BB}$), due to increased gravitational lensing. On the other hand, massive neutrinos help in reducing this smearing, because it decreases the gravitational lensing of the CMB photons, by suppressing the matter power spectrum in small scales, due to neutrinos having large thermal velocities which prevents them from clustering. Increasing the $\sum m_{\nu}$ parameter causes increasing suppression of matter power in the small scales \cite{Lesgourgues-Pastor}, which leads to decreasing gravitational lensing of the CMB photons. This leads to a strong positive correlation between $A_{\textrm{lens}}$ and $\sum m_{\nu}$, such as, to compensate for the increase in $A_{\textrm{lens}}$, the neutrino masses are also increased. The 1-D plots for $\sum m_{\nu}$ and $N_{\textrm{eff}}$ for selected datasets are given in figure~\ref{fig:9}. In this model, the Planck data is almost insensitive to neutrino masses $< 0.6$ eV. Our tightest bound of  $\sum m_{\nu} < 0.239$ eV (95\% C.L.)  again comes with Planck+BAO data. This bound, while weaker than the previous models we have discussed, is still close to the $\sum m_{\nu} < 0.23$ eV (95\% C.L.)  bound provided by Planck collaboration \cite{Ade:2015xua}, and still a large improvement compared to the ones presented in \cite{DiValentino:2017zyq}, which varied dark energy parameters in both in phantom and non-phantom range and had found a bound of $\sum m_{\nu}<$ 0.557 eV (95\% C.L.)  with Planck+BAO, demonstrating the large difference between phantom and non-phantom dark energies as far as neutrino masses are concerned. The preferred $N_{\textrm{eff}}$ values are also higher in NPDDE11+$A_{\textrm{lens}}$ compared to the previous cases. The addition of the HST data leads to even higher $N_{\textrm{eff}}$ which leads to the $N_{\textrm{eff}}=3.045$ value being disallowed even at 95\% C.L. with Planck+HST, for which the 68\% and 95\% limits are $N_{\textrm{eff}}=3.517_{-0.216}^{+0.196}$ and $N_{\textrm{eff}}=3.517_{-0.396}^{+0.424}$ respectively. This signifies the presence of tension between Planck and HST in this model, as it was in previous models.\\

The use of the $H071p6$ prior, again leads to lower values
of $N_{\textrm{eff}}$. In particular, with Planck + BAO + $H07106$, we get a bound of $N_{\textrm{eff}} = 3.329^{+0.207}_{-0.227}$ (68\%). Thus, $N_{\textrm{eff}} = 3.045$ is  not excluded at 95\% in this model, but excluded only at 68\%. \\

Another important change is the change in bounds on the optical depth to reionization, $\tau$. With Planck+BAO, the NPDDE11 model preferred a value of $\tau = 0.092\pm0.018$ (68\% C.L.), whereas this model prefers $\tau = 0.059^{+0.21}_{-0.22}$ (68\% C.L.), which is actually closer to the bound of $\tau = 0.055 \pm 0.009$ (68\% C.L.) given by Planck 2016 intermediate results \cite{Aghanim:2016yuo}. This was previously observed in \cite{DiValentino:2017zyq} which did the analysis with varying the dark energy parameters in both the phantom and non-phantom sector. This implies that the main effect is through the degeneracy between $\tau$ and $A_{\textrm{lens}}$ and has not much to do with dark energy. Again, while the NPDDE11+$r$ and NPDDE11 models failed to reconcile Planck with weak lensing measurements like CFHTLenS, the NPDDE11+$A_{\textrm{lens}}$ model prefers lower values of $\sigma_8$ and the agreement with CFHTLenS is considerable. This can be visualized in figure~\ref{fig:8}. This was also previously seen in \cite{DiValentino:2017zyq} and hence, again we can infer that this happens because of varying $A_{\textrm{lens}}$. The bounds on the dynamical dark energy parameters are however weaker than in the other two models. The cosmological constant is however compatible with the data even in this model. \\

\begin{figure}[tbp]
\centering % \begin{center}/\end{center} takes some additional vertical space
\includegraphics[width=.4963\linewidth]{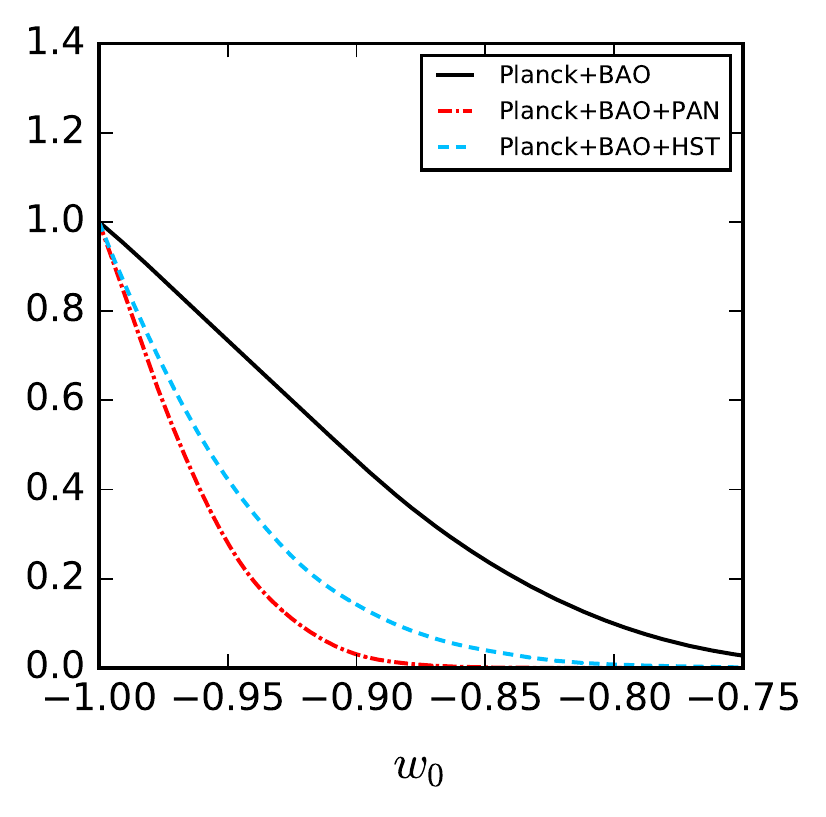}
\hfill
\includegraphics[width=.4963\linewidth]{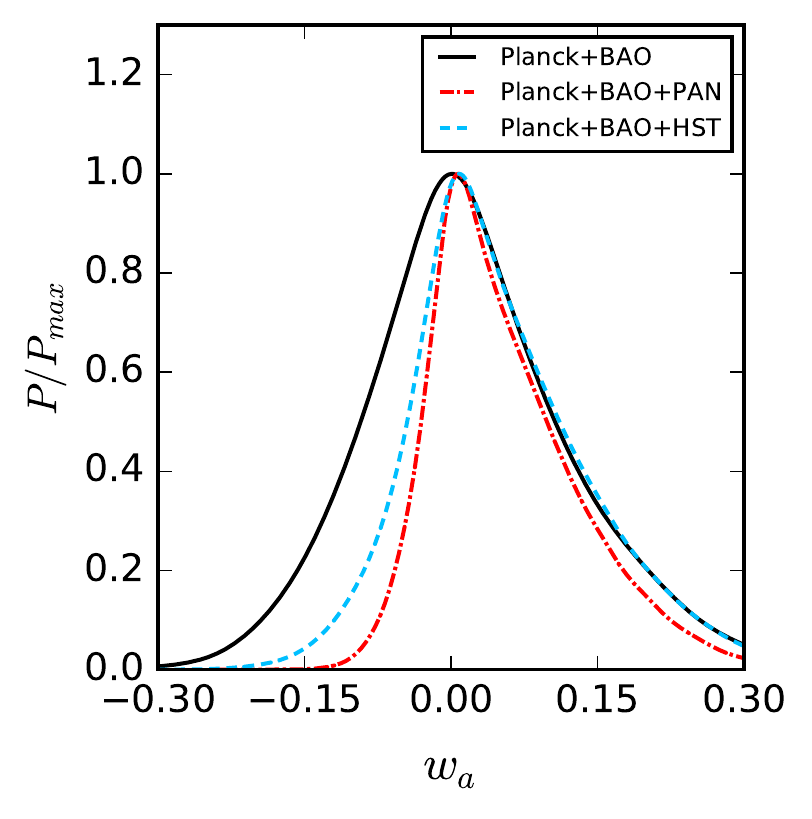}
% "\includegraphics" is very powerful; the graphicx package is already loaded
\caption{\label{fig:10}Comparison of 1-D marginalized posterior distributions for $w_0$ and $w_a$  for different data combinations in NPDDE11+$A_{\textrm{lens}}$.}
\end{figure}

\section{$\tau$ and $A_{\textrm{lens}}$: Implications for Planck 2018} 
\label{sec:6}
Both $\tau$ and $A_{\textrm{lens}}$ are correlated with $\sum m_{\nu}$, and with each other. In particular, when $A_{\textrm{lens}}$ is fixed, increase in $\sum m_{\nu}$ reduces smearing in the damping tail of the CMB power spectra, and it can be compensated by increasing $\tau$ \cite{Choudhury:2018byy,Vagnozzi:2017ovm}. Hence they have a positive correlation. On the other hand, increasing $A_{\textrm{lens}}$ increases the smearing of the damping tail, i.e., negative correlation with $\tau$. The value of $\tau$ has been significantly improved from Planck 2015 to Planck 2018. Thus we consider a bound on this optical depth to reionization, $\tau = 0.055\pm 0.009$, taken from \cite{Adam:2016hgk}, in which Planck collaboration removed previously unexplained systematic effects in the polarization data of the Planck HFI on large angular scales (low-$l$). We refer to this prior as $\tau 0p055$ hereafter. We use $\tau 0p055$ as a substitute for low-$l$ polarization data, and thus we discard the lowP data whenever we apply the $\tau 0p055$ prior, to avoid any double counting. This prior is very close to the bound, $\tau = 0.0544^{+0.0070}_{-0.0081}$ (68\%), obtained with Planck 2018 temperature and polarization data \cite{Aghanim:2018eyx}. Hence, imposition of $\tau = 0.055\pm 0.009$ would produce bounds on $\sum m_{\nu}$ that will be close to the bounds produced with Planck 2018 (instead of Planck 2015) in the models that we have considered. \\

We find that in the NPDDE11+$r$ model, with Planck + BK14 + BAO + $\tau 0p055$, we get $\sum m_{\nu} <$ 0.097 eV (95\%) (i.e. improvement over the  $\sum m_{\nu} <$0.123 eV limit as in \ref{table:2}, with Planck + BAO). This bound is actually lower than the $\sum m_{\nu}\simeq 0.1$ eV , i.e. minimum mass required for inverted mass hierarchy of neutrinos.  At the same time, in the NPDDE11 model, with Planck + BAO + $\tau 0p055$ we get $\sum m_{\nu} <$ 0.107 eV (95\%), which is also an improvement from the result: $\sum m_{\nu} <$ 0.126 eV (95\%) with Planck + BAO (see table \ref{table:4}. This happens, since in both of these models the mean value of $\tau$ hovers around 0.09-0.1. The $\tau 0p055$ prior partially breaks the degeneracy between $\tau$ and $\sum m_{\nu}$, and produces lower values of $\sum m_{\nu}$ by lowering the preferred $\tau$ values. \\

On the other hand, in the NPDDE11+$A_{\textrm{lens}}$ model with Planck + BAO + $\tau 0p055$, we found $\sum m_{\nu} < $ 0.237 eV, which is almost similar to the bound $\sum m_{\nu} < $ 0.239 eV  (95\%) with Planck + BAO (see table \ref{table:5}).  This happens since all the three parameters, $\tau$, $A_{\textrm{lens}}$, and $\sum m_{\nu}$ are varied together. Now, as the data prefers $A_{\textrm{lens}}$ values higher than the $\Lambda$CDM value in this model, the degeneracy between $A_{\textrm{lens}}$ and $\tau$ leads to a much lowered value of $\tau$, and thus the correlation between $\tau$ and $\sum m_{\nu}$ is already much smaller in this model, than the other two. Thus $\tau 0p055$ has little effect on the neutrino mass bounds in this model. \\

Also, we obtained limits of $A_{\textrm{lens}}$ in a $\Lambda\textrm{CDM} + A_{\textrm{lens}}$ model with Planck 2015 full temperature and polarization data. The value we got is $A_{\textrm{lens}} = 1.15^{+0.072}_{-0.082}$ (68\% C.L.). In the Planck 2018 Cosmological Parameters paper \cite{Aghanim:2018eyx}, for similar data and same model, given value of $A_{\textrm{lens}}$ is: $A_{\textrm{lens}} = 1.18\pm0.065$ (68\%) (see equation 36b). This shows that there is only a very small change in $A_{\textrm{lens}}$ from Planck 2015 to Planck 2018. Thus, it is likely that there will not be any considerable changes in the limits of other cosmological parameters with the Planck 2018 data, in the context of the value of $A_{\textrm{lens}}$.

\section{Summary}
\label{sec:5}

In this work we have studied three different extended cosmological scenarios with non-phantom dynamical dark energy (NPDDE) with a focus on constraining sum of neutrino masses. We have presented bounds on all the varying parameters in these extended scenarios and described the main effects we observed. In the first model, NPDDE11+$r$, we consider 12 parameters: the 6 $\Lambda \textrm{CDM}$ parameters, two dynamical dark energy parameters with CPL parametrization ($w_0$ and $w_a$) with hard priors to satisfy the non-phantom requirement, number of effective relativistic neutrino species at recombination ($N_{\textrm{eff}}$ and sum of neutrino masses ($\sum m_{\nu}$), and the running of the inflation spectral index ($n_{run}$) and the tensor-to-scalar ratio ($r_{0.05}$). We used different combinations of recent datasets including Planck 2015 temperature and polarization data, CMB B-mode spectrum data from BICEP2/Keck collaboration (BK14), BAO SDSS III BOSS DR12, MGS and 6dFS data, SNe Ia Pantheon sample (PAN), the HST prior ($H_0 = 73.24\pm 1.74$ km/sec/Mpc (68\% C.L.)). We found that CMB only data is not very effective in constraining the cosmological parameters. The 1$\sigma$ spreads for the parameters were however increased in this model compared to $\Lambda \textrm{CDM}$ due to the doubling of number of parameters. Our best bound on neutrino masses in this model came from Planck+BK14+BAO: $\sum m_{\nu} < 0.123$ eV (95\% C.L.)  which is a strong bound close to the minimum mass of $\simeq$ 0.1 eV (95\% C.L.)  required for inverted hierarchy of neutrino masses and is stronger than a bound of $\sum m_{\nu} <$ 0.158 eV (95\% C.L.)  obtained in $\Lambda \textrm{CDM}+\sum m_{\nu}$ with Planck+BAO \cite{Choudhury:2018byy} (see also \cite{Vagnozzi:2018jhn} for a similar conclusion in a smaller parameter space). We also found that inclusion of the HST prior leads to a preference for dark radiation at 68\% C.L. but not at 95\%, while without the HST prior the data is completely consistent with the standard value of $N_{\textrm{eff}} = 3.045$. Although this is driven by the more than 3$\sigma$ tension present between Planck and HST regarding the value of $H_0$ and should be interpreted cautiously. This model did not improve the $\sigma_8$ tension present in the $\sigma_8-\Omega_m$ plane between Planck and CFHTLenS. The Pantheon sample improved the bounds on the dark energy parameters. All combinations of data are also compatible with a cosmological constant ($w_0=-1, w_a = 0$). However, this is mostly because we are restricting the parameter space to $w(z)\geq -1$ and \cite{DiValentino:2017zyq} had found that the data mostly prefers the phantom region in such an extended parameter space when both phantom and non-phantom regions are allowed.\\

We tested the stability of these results in a lower parameter space (model:NPDDE11) where we turned off the tensor perturbations and also did not use the BK14 data. We found that the general conclusions made for NPDDE11+$r$ were also true in this model. The tightest bound of $\sum m_{\nu} <$ 0.126 eV (95\% C.L.) in this model also came from Planck + BAO. \\

Finally we studied the NPDDE11+$A_{\textrm{lens}}$ model where we also varied the lensing amplitude. We found that except when Planck lensing data is included, the $A_{\textrm{lens}} = 1$ value predicted by $\Lambda \textrm{CDM}$ was rejected at more than 95\% C.L. by the datasets. Due to this, the $\sum m_{\nu}$ bounds also worsened with our best result in this model:$\sum m_{\nu} < 0.239$ eV (95\% C.L.)  coming from Planck+BAO again. This result is, however, still close to the $\sum m_{\nu} < 0.23$ eV (95\% C.L.)  bound by Planck collaboration \cite{Ade:2015xua}, showing that the cosmological data is effective in constraining neutrino masses in a cosmology with NPDDE. The HST prior also preferred a dark radiation component but this time also at 95\% C.L. level, as this model also prefers higher values of $N_{\textrm{eff}}$. On the other hand, we found that this model helps relieve the $\sigma_8$ tension between Planck and CFHTLenS considerably.	 \\  

The recent Planck 2018 results \cite{Aghanim:2018eyx} put the bound of $\sum m_{\nu}<$ 0.13 eV (95\% C.L.)  in $\Lambda \textrm{CDM}+\sum m_{\nu}$ with Planck+BAO. Thus, the aggressive bound of $\sum m_{\nu}<$ 0.123 eV (95\% C.L.) (Planck + BK14 + BAO) is still stronger than this bound by Planck 2018 and hence, our results are very much relevant albeit the analysis is with Planck 2015 dataset. In fact, when we use the following Gaussian prior on optical depth to reionization: $\tau = 0.055\pm0.009$ from 2016 Planck intermediate results, and discard the low-$l$ polarization data, the bound on neutrino masses improves to $\sum m_{\nu} <$ 0.097 eV (95\%), which is less than the 0.1 eV mass sum required for inverted hierarchy of active neutrino masses. \\

While we have used the CPL parameterization in our paper, it is not the only parameterization that can be used for non-phantom dark energy. Any change in parameterization can lead to change in bounds obtained on the sum of neutrino masses. For instance, if we set the $w_a$ parameter to zero, i.e., if we consider only a simple $w(z) = w_0$ parameterization, we find that bounds on $\sum m_{\nu}$ relax slightly. In the NPDDE11 model, with $w_a = 0$ and $w(z)=w_0$, and using Planck + BAO data, we found $\sum m_{\nu} <$ 0.141 eV (95\%), instead of $\sum m_{\nu} <$ 0.126 eV (95\%) when we vary both $w_0$ and $w_a$. In the NPDDE11+$A_{\textrm{lens}}$ model also, with $w_a = 0$ and $w(z)=w_0$, and using Planck + BAO, we obtained $\sum m_{\nu} <$ 0.261 eV (95\%), instead of $\sum m_{\nu} <$ 0.239 eV (95\%). Some other parameterizations that can be considered include Logarithmic parameterization \cite{Efstathiou:1999tm} ($w(a) = w_0 - w_aln(a)$), Jassal-Bagla-Padmanabhan (JBP) parameterization \cite{Jassal:2004ej} ($w(a) = w_0 + w_a a(1-a)$ etc. Analysis involving these parameterizations is beyond the scope of our work in this paper. However, we would like to point the reader to \cite{Yang:2017amu}, where the authors found similar limits, with CPL and Logarithmic parameterizations, on $\sum m_{\nu}$ for the case of degenerate hierarchy. However, in case of JBP, bound on $\sum m_{\nu}$ was found to be significantly stronger. While \cite{Yang:2017amu} does not discard the phantom region, it is possible that results from analyses with only non-phantom dark energy will also vary depending on the parameterization used, as far as neutrino masses are concerned. \\
 
We would like to add a final remark that we have obtained the bounds while taking the datasets at face value. However unresolved systematics present in the dataset could have affected our results and conclusions. For instance the tension between Planck and HST prior can be due to a dark radiation species, but can also be due to systematics present in both the datasets. Thus there is still a lot to learn about robustness of datasets and also about dynamics of dark energy.

\acknowledgments
The authors sincerely thank the anonymous referee for the thoughtful comments and efforts towards improving our manuscript. SRC thanks the cluster computing facility at HRI (\url{http://www.hri.res.in/cluster/}) and the Department of Atomic Energy (DAE) Neutrino Project of HRI. AN thanks ISI Kolkata for financial support through Senior Research Fellowship. The authors also thank Archita Bhattacharyya for valuable help. This project has received funding from the European Union's Horizon 2020 research and innovation programme InvisiblesPlus RISE under the Marie Sklodowska-Curie grant agreement No 690575. This project has received funding from the European Union's Horizon 2020 research and innovation programme Elusives ITN under the Marie Sklodowska-Curie grant agreement No 674896. 

\bibliography{paper}
\bibliographystyle{jhep}

\end{document}